\documentclass[%
 reprint,
 amsmath,amssymb,
 aps,
]{revtex4-2}
\usepackage[utf8]{inputenc}
\usepackage[T1]{fontenc}
\usepackage{subfloat}
\usepackage{graphicx}
\usepackage{subfig}
\usepackage{url}

\begin{document}

\title{Quantifying relation between mobility patterns and socioeconomic status of dockless sharing-bike users}

\author{Tianli Gao, Zikun Xu$^*$, Chenxin Liu, Yu Yang, Fan Shang, Ruiqi Li}
\thanks{Corresponding author: lir@buct.edu.cn (R. Li) \\ 
Corresponding author: rammy\_xu@foxmail.com (Z. Xu)}
\affiliation{UrbanNet Lab, College of Information Science and Technology, Beijing University of Chemical Technology, Beijing 100029, China}





\begin{abstract}
Bikes are among the healthiest, greenest, and most affordable means of transportation for a better future city, but mobility patterns of riders with different income were rarely studied due to limitation on collecting data. Newly emergent dockless bike-sharing platforms that record detailed information regarding each trip provide us a unique opportunity. Attribute to its better usage flexibility and accessibility, dockless bike-sharing platforms are booming over the past a few years worldwide and reviving the riding fashion in cities. In this work, by exploiting massive riding records in two megacities from a dockless bike-sharing platform, we reveal that individual mobility patterns, including radius of gyration and average travel distance, are similar among users with different income, which indicates that human beings all follow similar physical rules. However, collective mobility patterns, including average range and diversity of visitation, and commuting directions, all exhibit different behaviors and spatial patterns across income categories. Hotspot locations that attract more cycling activities are quite different over groups, and locations where users reside are of a low user ratio for both higher and lower income groups. Lower income groups are inclined to visit less flourishing locations, and commute towards the direction to the city center in both cities, and of a smaller mobility diversity in Beijing but a larger diversity in Shanghai. In addition, differences on mobility patterns among socioeconomic categories are more evident in Beijing than in Shanghai. Our findings would be helpful on designing better promotion strategies for dockless bike-sharing platforms and towards the transition to a more sustainable green transportation.
\end{abstract}

\maketitle

\thispagestyle{empty}

\section{Introduction}

Bike once was one of the main and fancy transportation means many decades ago especially when the spatial size of cities is relatively small. Later, almost world-widely, with the booming of the car industry since the 20th-century \cite{jacobs1961death}, most modern cities have followed a car-centric development along with rapid urbanization \cite{li2017simple,website:UNprospects,xu2019cross,gerland2014world,li2021assessing}.  
Although bikes soon became affordable to the great majority of people, the usefulness and dominant position of the bike as a mean for commuting is challenged, and attitudes of the public towards biking are also changing \cite{underwood2014teens}. 
However, the flooding of cars and car-orientated transportation systems are responsible for a variety of urban illnesses, including traffic congestion, air pollution, energy deficiency, and deterioration of human health \cite{WHO2002physically, demaio2009bike,jappinen2013modelling}. 
Though with an aim to provide a solution to sustainable mobility, car-sharing platforms have intensified traffic congestion on both intensity and duration since their launching \cite{diao2021impacts}. 
In the past few decades, with growing concerns over global warming and rapid urban sprawl, numerous efforts have been devoted to promoting bike-sharing in cities as a viable, greener, more resilient, and healthier mobility solution towards the transition to sustainable transportation \cite{dill2003bicycle,jappinen2013modelling,hull2014bicycle}. As biking can reduce carbon emissions and improve public health due to its usage flexibility, saving of parking space and fossil energy by replacing short-distance motorized trips, increasing fitness, and reducing the stress of riders from cycling activities \cite{xu2019unravelBIKE,jiang2020dockless,2019bikelife,cheng2021role}. During the COVID-19 pandemic, bike-sharing systems are proved to be more resilient \cite{teixeira2020link} and safer to move around for essential needs in urban area with a high population density \cite{li2018effect}, since biking allows for greater social distancing than other means of public transportation \cite{jiang2020dockless}. Biking is also expected to be more frequent in cities after the pandemic \cite{2020meninoSurvey}. 

With technology advances in IoT (Internet of Things) and mobile payment, dockless bike-sharing platforms emerged worldwide, such as Mobike, Ofo Bike, DiDi Bike, LimeBike, Spin, and Ford GoBike. Compared to ordinary docked public sharing bikes that need to be returned to docks in fixed places, dockless ones are free from such restrictions and give better accessibility and flexibility to users \cite{li2021gravity,luo2020revealing,li2022emergence}. Users can return, simply park and lock, a bike anywhere suitable for a bike near the destination. 
So far, dockless sharing bikes are appearing in over 360 cities with 47 million trips per day \cite{jiang2020dockless}.

Given the potential benefits of cycling activities on both the health of individuals and the whole transportation system and the fact that bikes are quite affordable, understanding the relationship between mobility patterns of dockless sharing bike users and their socioeconomic status could be crucial for better urban design \cite{olmos2020data}, the development of the green economy and sharing economy \cite{shaheen2016mobility}.  
Previous studies suggest that income can have a strong influence on mobility patterns \cite{Xu2018HumanMA,Barbosa2021hundred,lenormand2015influence,pan2011fortune}, will this be the case for biking? 
However, most studies are based on mobile phone data \cite{Xu2018HumanMA,pan2011fortune} and credit card transactions \cite{lenormand2015influence} that reflect a mixture of multiple travel modes (e.g., biking, driving, flying, and even walking). In comparison, few work is dedicated to cycling via dockless sharing bikes. 

In this work, by exploiting massive riding records from Mobike, a dockless bike-sharing system, and house price data as a proxy of income, we investigate the relation between mobility patterns of riding behaviors and socioeconomic status in Beijing and Shanghai. We first observe that the user ratio has a declining trend along with the increase of income, and both richest and poorest groups are of a relatively lower user ratio. Users from higher and lower income groups are of a more uneven adoption rate and reside more concentrated in certain places in both cities, which suggest greater promotion potentials for locations with lower adoption rate for these two groups. 
We then reveal that individual mobility patterns, such as radius of gyration and travel distance, are similar among people with different income, which indicates that human beings all follow similar physical rules. However, collective mobility patterns, including extent of dispersion, diversity of visitation, and spatial patterns of cycling activities, all exhibit different behaviors across different income categories. 
The most visited locations (i.e., hotspots) of each group are quite different and spatially separated from each other in Beijing, and relatively similar in Shanghai. Lower income groups are inclined to visit less flourishing locations with a different extend, and commute towards the direction to the city center in both cities. 
The average range of cycling activities of lower-income groups in Beijing are larger which suggest that they have fewer visitations to locations near the central regions. In comparison, the average range of cycling activities of users from different income categories in Shanghai are more similar. We also observe that cycling activities of the highest and lower-income groups in Beijing are more concentrated as indicated by a lower diversity, while in Shanghai, an opposite pattern is observed and the diversity is larger than in Beijing. In addition, differences on mobility patterns among socioeconomic categories are more evident in Beijing than in Shanghai.

The paper is organized as follows. In Section \ref{sec:relatedwork}, we review related works on the relationship between mobility patterns and the socioeconomic status. In Section \ref{sec:approach}, we introduce methods used in this work. In Section \ref{sec:setup}, we describe datasets involved in this work and the data processing procedures, and related basic analyses. In Section \ref{sec:results}, we demonstrate results on the relationship between mobility patterns and the socioeconomic status at both the individual and collective level, and a discussion is given in Section \ref{sec:conclusion}.

\section{Related Works\label{sec:relatedwork}}
Attribute to accumulation of various types of geo-located data with high spatio-temporal resolution, e.g., cell detailed records \cite{alexander2015origin,dong2016population,jiang2013review,xu2017clearer}, GPS \cite{zheng2011learning,pappalardo2015returners}, credit card transactions \cite{lenormand2015influence}, mobility patterns have been extensively studied over past decades \cite{barbosa2018human} and its relation with socioeconomic status of individuals. 


By exploiting massive cellphone data, Pan et al. discovered that richer people with abundant resources tend to explore more places (e.g., shops, restaurants) out of curiosity and social motivations and this is more evident in the wealthiest groups \cite{pan2011fortune}.
Xu et al.  
revealed that individual mobility patterns (e.g., travel distance \cite{brockmann2006scaling,gonzalez2008understanding}, radius of gyration \cite{gonzalez2008understanding}, exploration and revisitation dynamics \cite{song2010modelling,pan2011fortune}, and activity entropy \cite{song2010limits}) are of little difference over people with different income \cite{Xu2018HumanMA}. However, at the collective level, the travel range, indicated by the radius of gyration, of richer people is shorter in Singapore but longer in Boston, and vice versa for people with less fortune \cite{Xu2018HumanMA}. 
Such a finding echos some earlier works that show the advantaged class tends to have a higher travel distance in some cities \cite{Poston1972metropolitan}, but a shorter distance in other cities (e.g., in Europe \cite{AGUILERA2009commuting}). 
This difference could be explained by the spatial distribution of neighborhoods in cities. 
More than a half century ago, Mills assumed that people with fortune tend to live in the suburban area based on studying traditional North American cities \cite{mills1967aggregative}. However, as time shifts, the rich begin to move back to the center in many cities. 
After analysing the 50 largest cities in the United States, Barbosa et al. find that there are two typical clusters: in one group, mobility patterns are strongly influenced by socioeconomic status; while, in the other, they two are almost irrelevant \cite{Barbosa2021hundred}. 
However, in Brazil, after grouping the 50 largest cities into two clusters,  both show a correlation between mobility patterns and income levels, only that the correlation is stronger in one group than the other  \cite{Barbosa2021hundred}. 
Accessibility of public transportation and availability of the service might be responsible for the observed division. 

However, most works are not focusing on specific travel mode of mobility, for example, mobility patterns inferred from cellphone data can be a mixture of multiple travel modes, including driving, biking, or even walking. 
In comparisons, there are relatively fewer works dedicating to the biking behaviors due to limitations on data accessibility, let alone on newly emergent dockless sharing bikes.

\section{Experimental Setup\label{sec:setup}}
In this section, we give a description of datasets involved during the study, demonstrate the process to extract related information from the raw data, and offer a brief analysis of the data.

\subsection{Data Description}
In this work, the dockless bike-sharing dataset is obtained from the Mobike platform. It contains over 3.1 million records in Beijing and 1.02 million records in Shanghai. Each record contains ``the order ID, the user ID, the bike ID, the departure and arrival locations with their timestamps''. In Beijing, there were 348 thousand users with trip records from May 10th to 24th in 2017. 
In Shanghai, 17.7 thousand users are recorded from Aug. 1st to Sep. 1st in 2016. 

To obtain a proxy of income of users residing in different locations, we get house price data from a real estate agency that wants to remain anonymous.  
There are 90 thousand records in Beijing and 60 thousand records in Shanghai. Each record contains the house's total price (referred as ``house price'' in this work), price per square meter (referred as ``unit price'' hereafter), floor space in square meter, and the longitude and latitude. The average house price of a location refers to the average of house price of all apartments (or houses, very occasionally) there. 

\begin{figure*}[!htbp]\centering
\includegraphics[width=\linewidth]{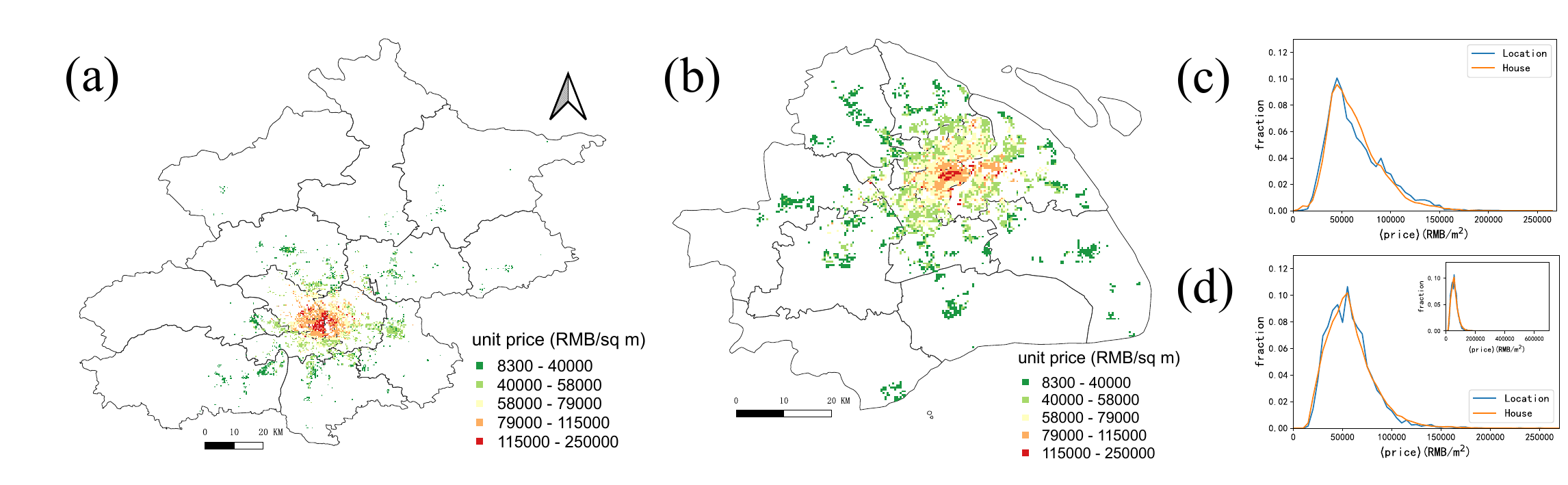}
\caption{The spatial distribution of unit house price per square meter (i.e., unit price) in RMB in Beijing (a) and Shanghai (b), and the distribution of unit price of apartments (blue lines) and distribution of average unit price of locations (orange lines) in Beijing (c) and Shanghai (d). Inset of (d) is the full distribution in Shanghai.}
\label{fig:spatial_price}
\end{figure*}

We rasterize the urban space into locations with $500\textit{m}\times 500\textit{m}$, thus, in this work, a location refers to such a place with an edge length of 500 meters. And we visualize the spatial distribution of unit price of locations (see Fig. \ref{fig:spatial_price}), where a clear spatial gradient present in both Beijing and Shanghai. The unit price is highest in the central area, and decrease when goes further away from the city center. Here, the unit price of a certain location is the average of unit price of all apartments there. The distribution of unit price are of similar shape in Beijing and Shanghai (see Fig. \ref{fig:spatial_price}c,d), but there are more extreme outliers in Shanghai (see the inset of Fig. \ref{fig:spatial_price}d). 
The distribution of unit price (i.e., price in square meters) of locations have a very similar to the distribution of unit price of all apartments in both cities (see Fig. \ref{fig:spatial_price}). This indicate that such a coarse-graining process does not affect the resolution of the data.

Besides, to verify if a user really live in a certain location, we double check the inferred home location with point-of-interests (POIs) obtained from the AMAP API (\url{https://lbs.amap.com}) of that location. 
If there is no POI with a type of ``Commercial House'' returned for that location, then we will not associate the place as the home to any user.

\subsection{Data Processing}
\subsubsection{Data Filtering}
\begin{figure}[!htbp] \centering
\subfloat[]{\includegraphics[width=0.5\linewidth]{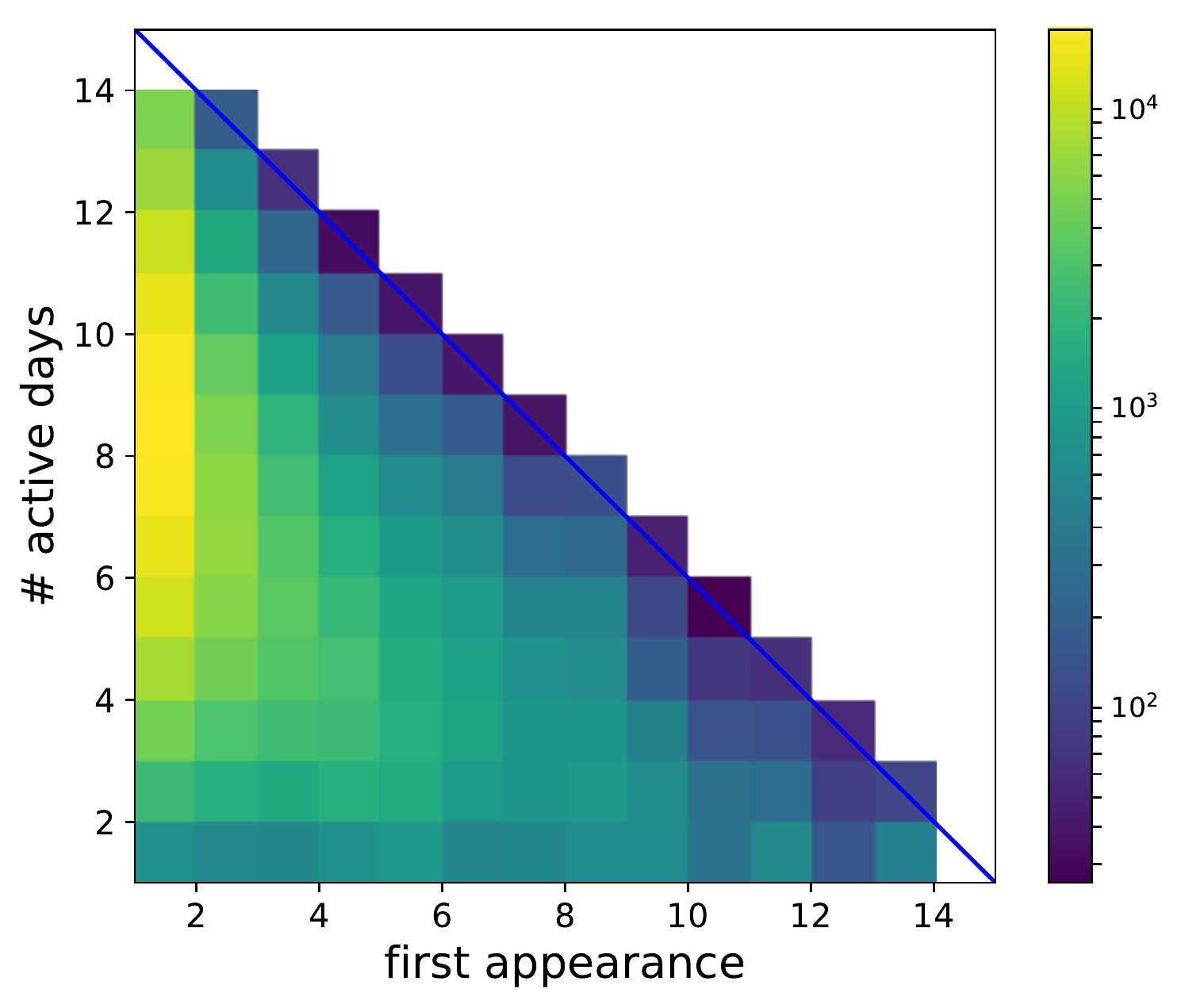}%
\label{fig:first_use_beijing}}
\hfil
\subfloat[]{\includegraphics[width=0.5\linewidth]{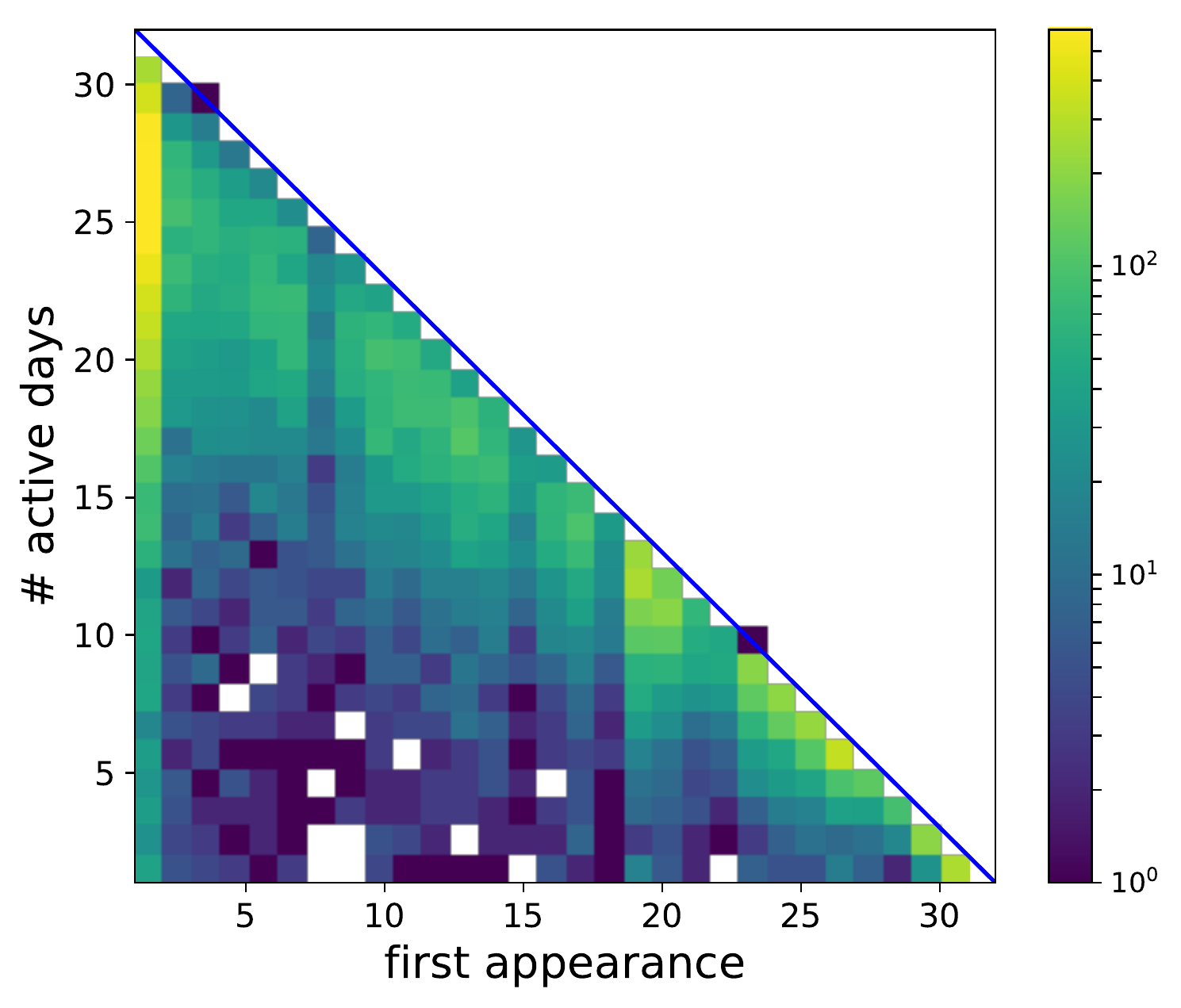}%
\label{fig:first_use_shanghai}}
\caption{The heat map on the number of users with certain first appearance date (indexed from the first day of the datasets) and the number of days with cycling activities in Beijing (a) and Shanghai (b). The blue line indicates the boundary where the sum of cycling days and first appearance day equal to the length of the dataset.}
\label{fig:first_use}
\end{figure}

\begin{figure}[!htbp]\centering
\subfloat[]{\includegraphics[width=0.5\linewidth]{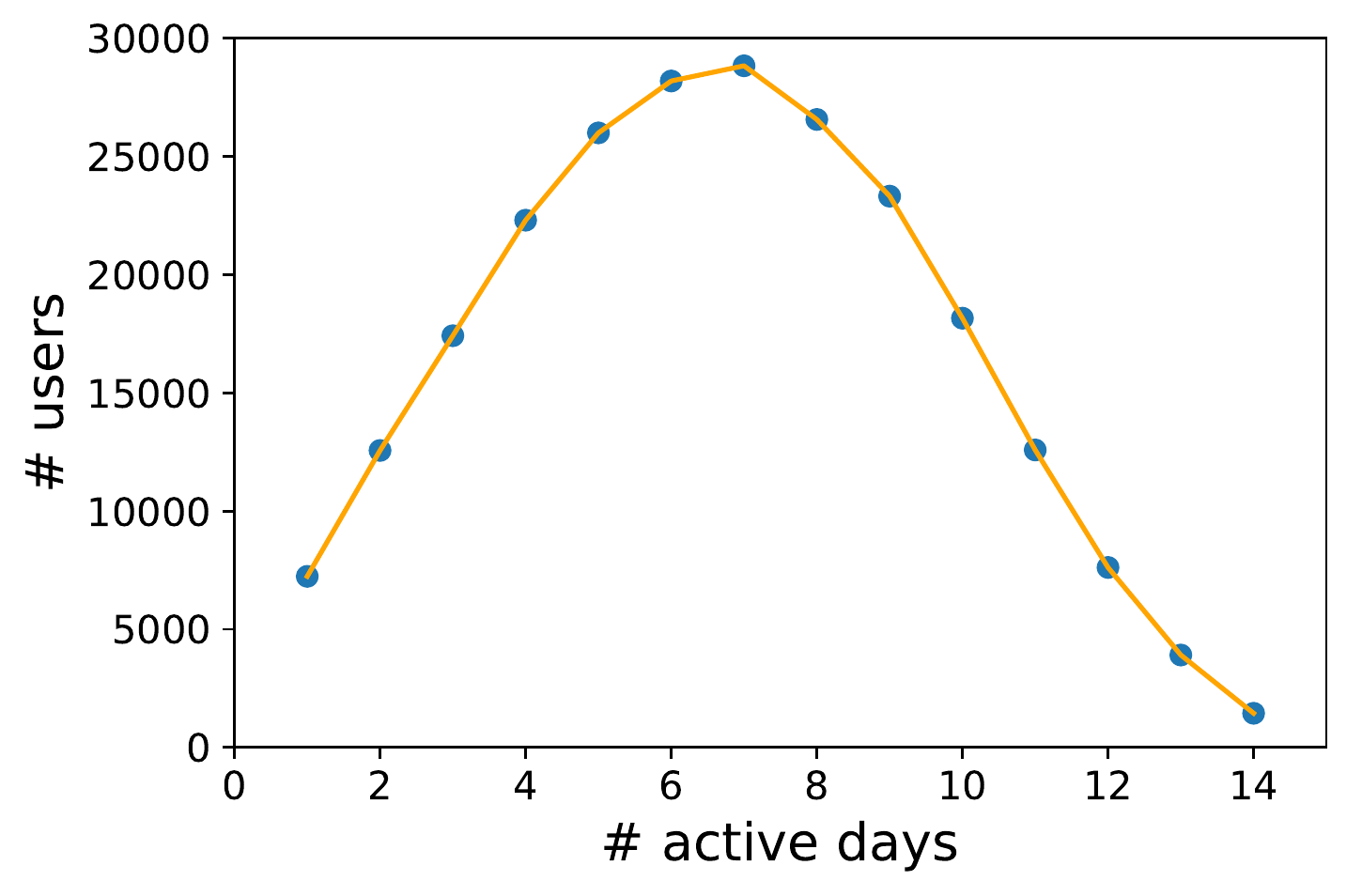}%
\label{active_day_beijing}}
\hfil
\subfloat[]{\includegraphics[width=0.5\linewidth]{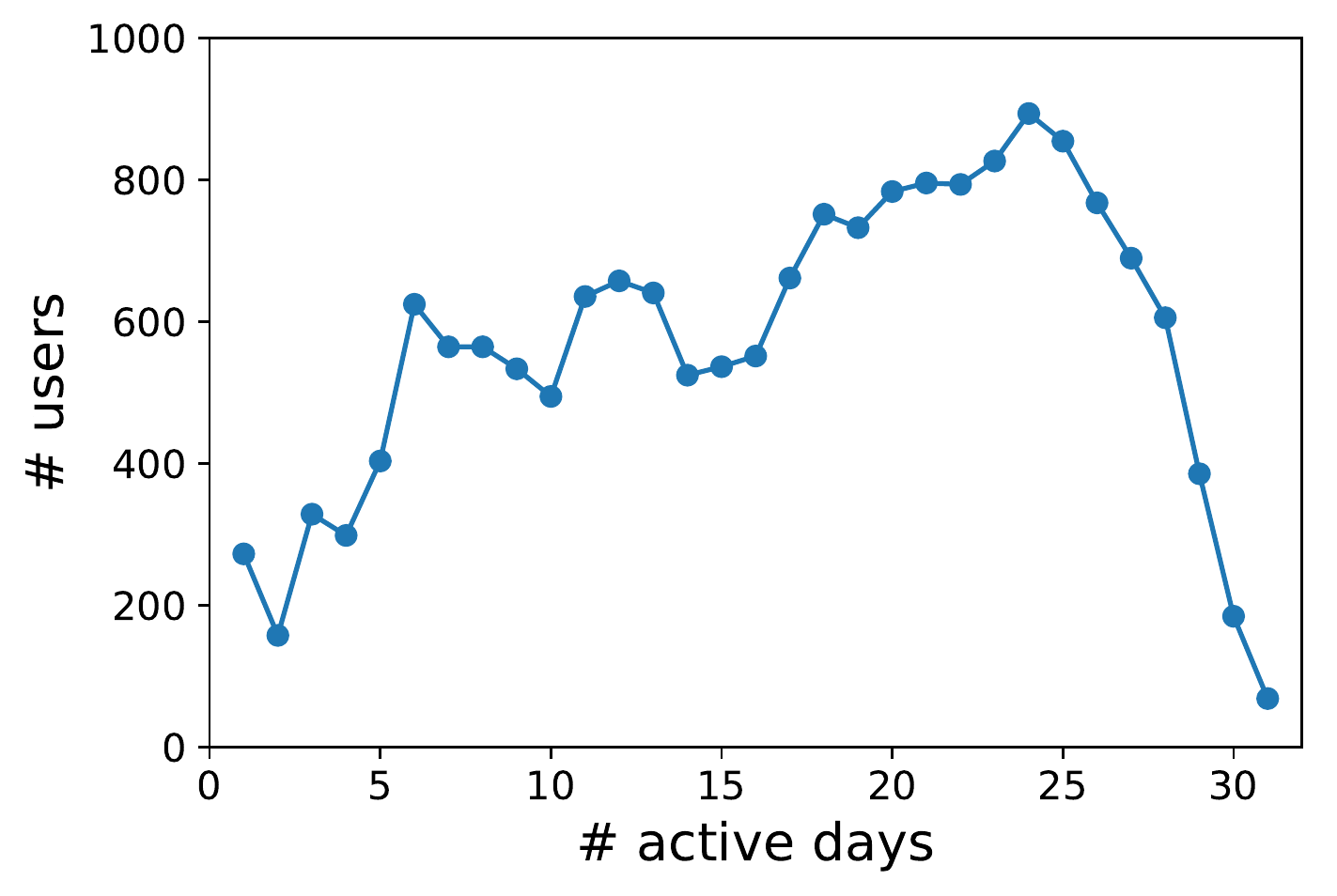}%
\label{active_day_shanghai}}
\caption{The frequency of the number of cycling days of users in Beijing (a) and Shanghai (b). 
}
\label{active_day}
\end{figure}

For the raw data, we first do some simple filtering: we discard  records not located within the boundary of the city and also discard the ones with a riding duration longer than one day (which might be some bikes left unlocked after their initial order) or less than one minute (which might be due to unsatisfied tryouts). In the Beijing dataset, 0.120 million out of 3.2 million records are discarded, most of which are the ones not located in the city; while in the Shanghai dataset, around 200 noisy records are discarded. 
As we need to identify the home and work locations of a user in this study, 
We further filter out all users who do not have two or more trips in any day. 
In Beijing, there are 2.77 million records remaining for 0.236 million users.
In Shanghai, 1.02 million records are available after the filter for 17.6 thousand users. 
Note that users might have different number of days with cycling activities (see Fig. \ref{fig:first_use_beijing}). 
Most active users appeared in the first day of in both datasets, and use dockless sharing bikes for around seven days in Beijing (see Fig. \ref{fig:first_use_beijing} and Fig. \ref{active_day_beijing}), and twenty-four days in Shanghai (see Fig. \ref{fig:first_use_shanghai} and Fig. \ref{active_day_shanghai}).  
Such a criterion can avoid common problems of a fixed threshold (such as the number of trips), as some users might only appear quite late during the period of the dataset, but they can be quite active and these data should not be filtered. For example, there are many users appeared quite late in the Shanghai dataset, but they are cycling for almost all of the remaining days in that month (see yellow regions on the right side of Fig. \ref{fig:first_use_shanghai} near the diagonal line). 

\begin{figure}[!htbp] \centering
\subfloat[]{\includegraphics[width=0.5\linewidth]{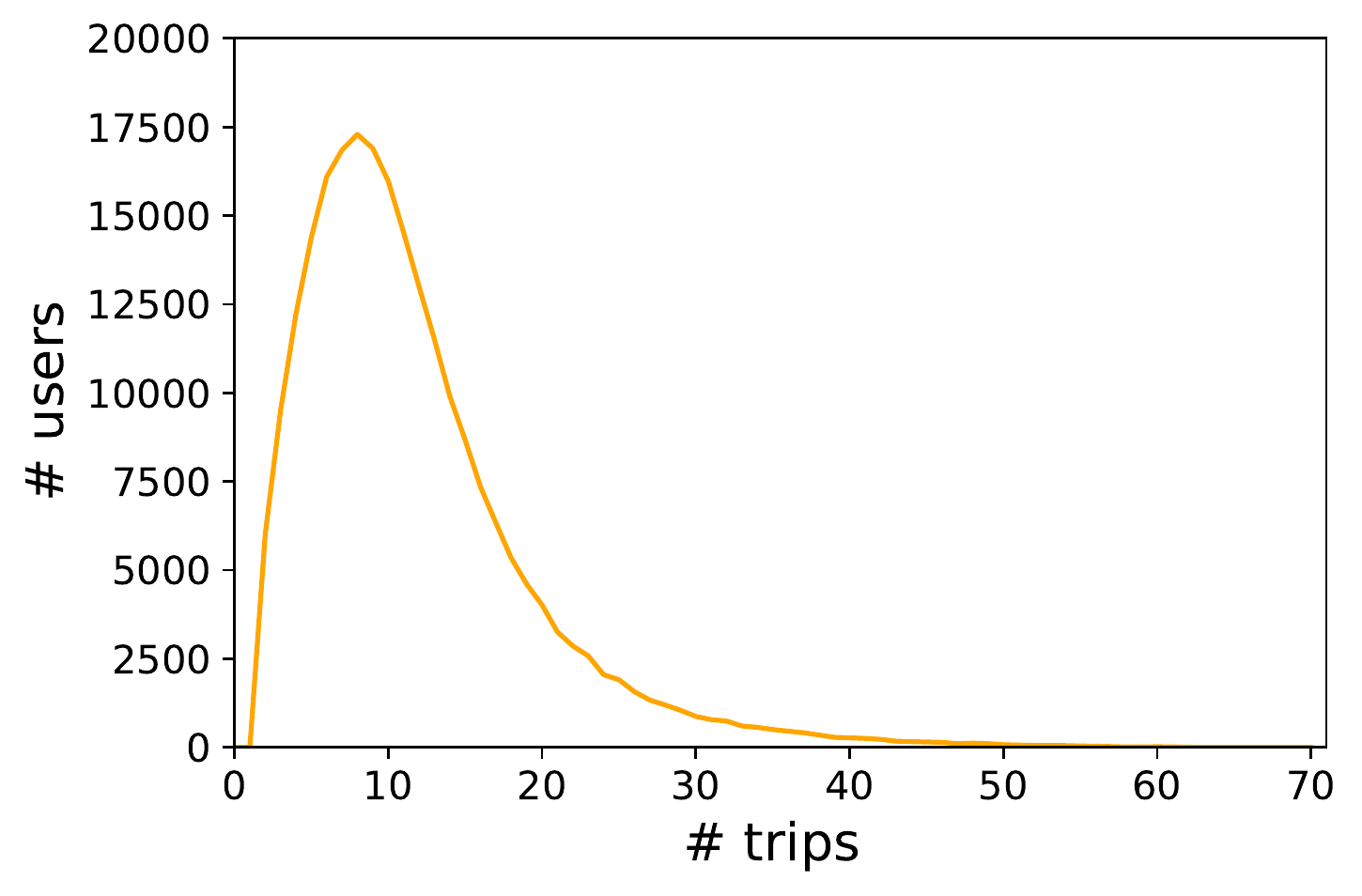}%
\label{travel_time_beijing}}
\hfil
\subfloat[]{\includegraphics[width=0.5\linewidth]{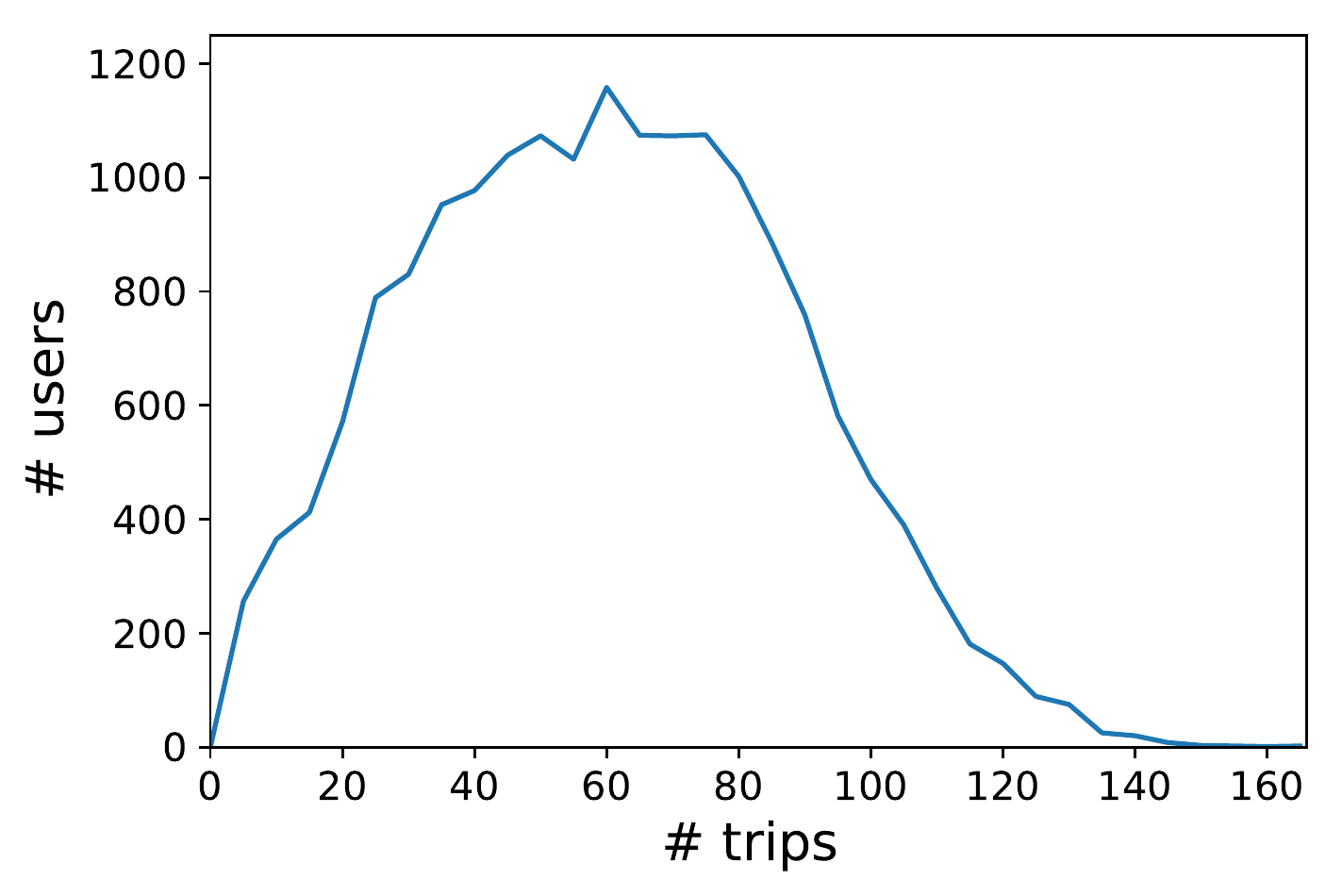}%
\label{travel_time_shanghai}}
\caption{The frequency of the number of trips of active users in Beijing (a) and Shanghai (b).} 
\label{travel_time}
\end{figure}

Besides, we find the distributions of cycling trips of active users in Beijing and Shanghai are relatively similar (see Fig. \ref{travel_time}) only with Beijing's distribution with a longer tail. 

\begin{figure*}[!t] \centering
\subfloat[]{\includegraphics[width=0.5\linewidth]{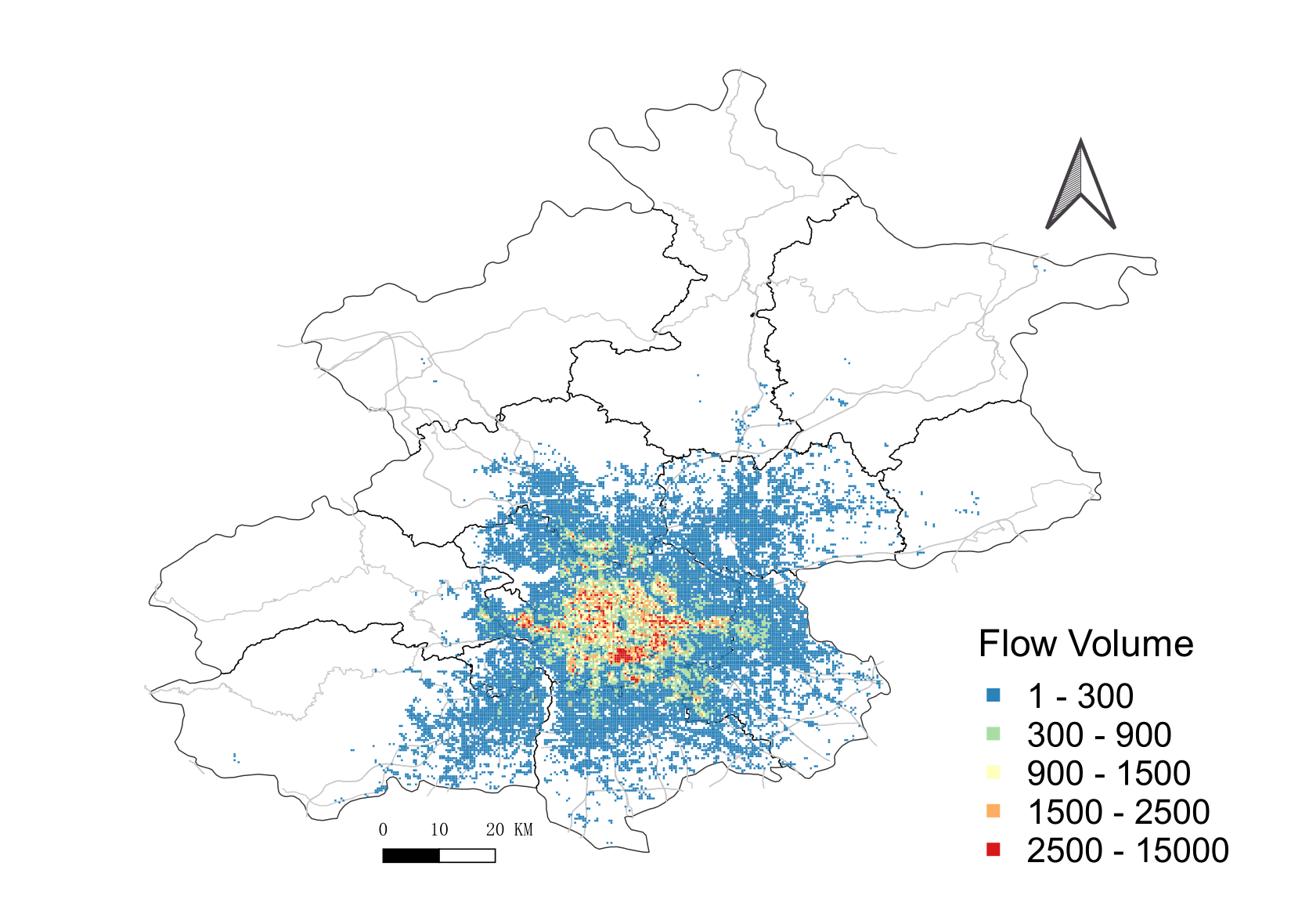}
\label{fig:spatial_flow_beijing}}
\hfil
\subfloat[]{\includegraphics[width=0.5\linewidth]{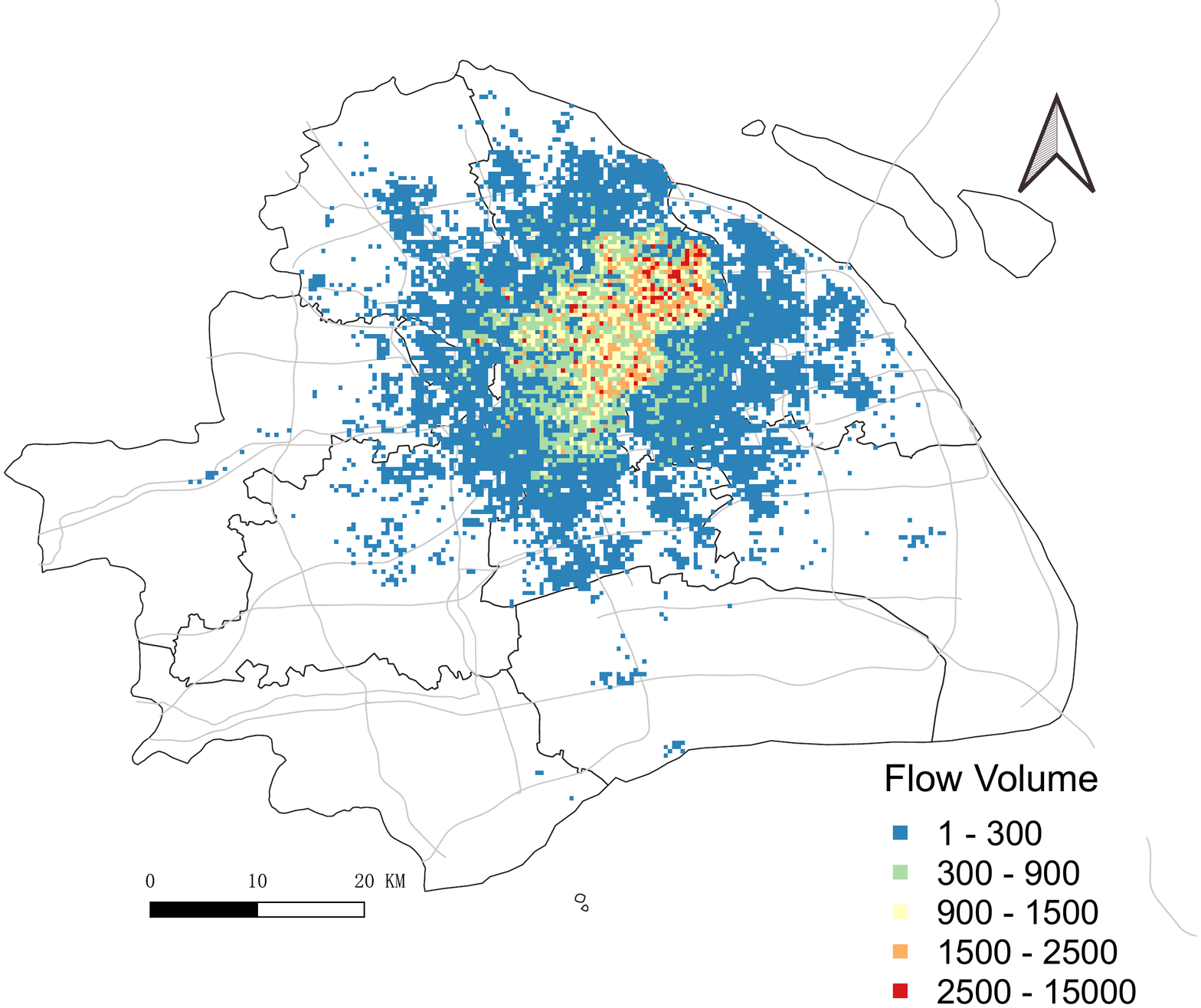}%
\label{fig:spatial_flow_shanghai}}
\caption{The spatial distribution of flow volume at the location level in Beijing (a) and Shanghai (b).}
\label{fig:spatial_flow}
\end{figure*}

We then aggregate all remaining trips to the location level, and visualize the spatial distribution of the total number of trips either departing from or entering locations $T_i$ (referred as ``flow volume'' hereafter, see Fig. \ref{fig:spatial_flow}). 
It also exhibits a spatial gradient, which is 
similar to the declining of house price, but the hotspot locations that are of highest values are different from the ones in the landscape of house price (see Fig. \ref{fig:spatial_flow} and Fig. \ref{fig:spatial_price}). In Beijing, locations with the highest flow volume are surrounding the city center where house price peaks; and in Shanghai, these locations slightly moving to north-eastern direction. This shows that the locations with the highest house price are not having the largest biking flow. 

\subsubsection{Obtaining Proxy for Income Level}
Due to data access limitations, the income data of residents is not available, which is a common problem encountered by most studies \cite{xu2019quantifying,xu2019unraveling,xu2017clearer,alfeo2019assessing}. 
A common practice is to use average house price of locations where users reside as a proxy for their economic status \cite{xu2019quantifying,xu2019unraveling,xu2017clearer,alfeo2019assessing}. 
Some evidence also shows that there is a clear linear relation between the average housing price and average monthly income at the planning area resolution in Singapore \cite{Xu2018HumanMA}. 
We also use average house price of locations as the proxy for the average income there in this work. 

\begin{figure}[!b] \centering
\subfloat[]{\includegraphics[width=0.5\linewidth]{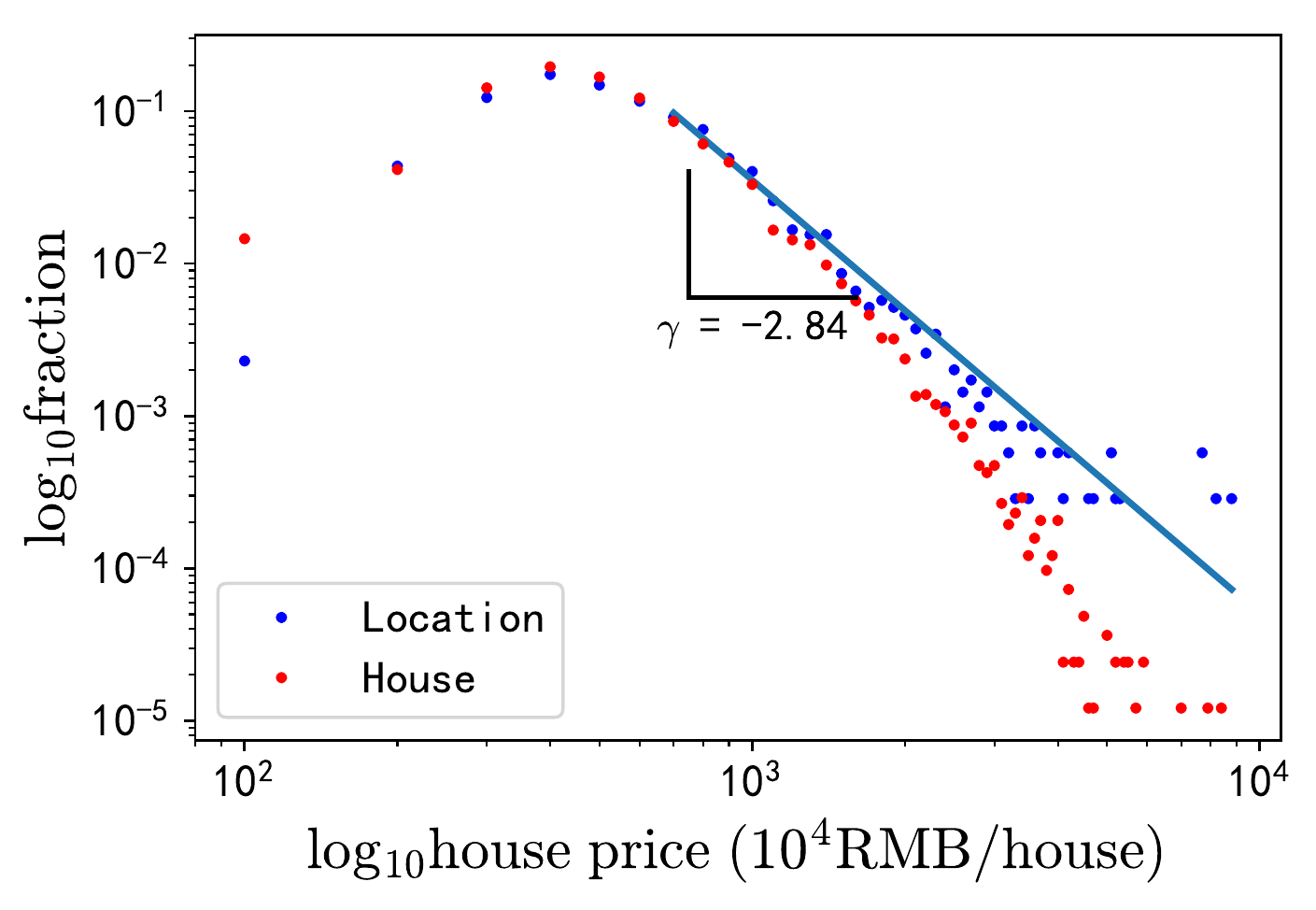}%
\label{fig:housing_total_price_freq_beijing}}
\hfil
\subfloat[]{\includegraphics[width=0.5\linewidth]{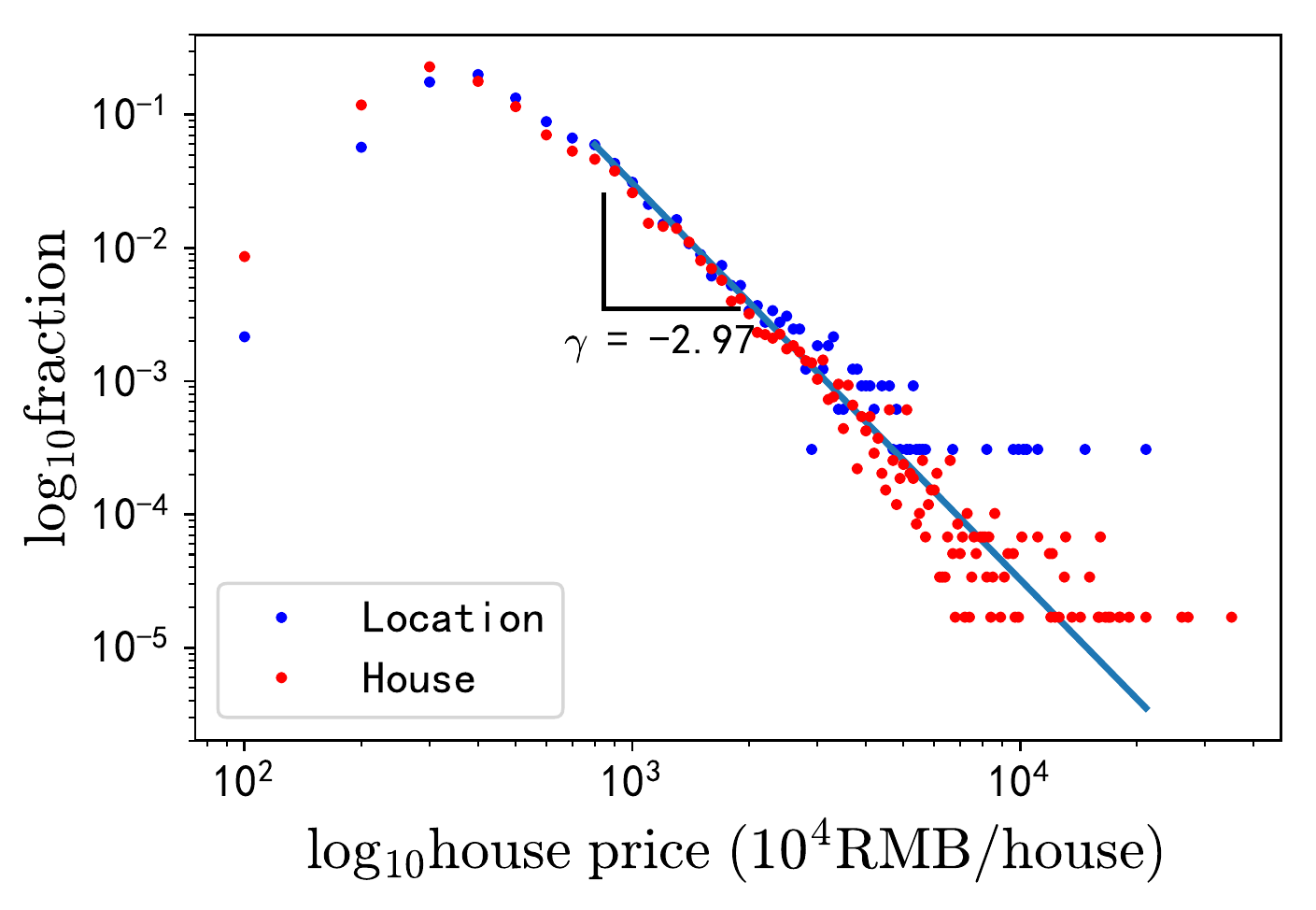}%
\label{fig:housing_total_price_freq_shanghai}}
\caption{The distribution of house price of apartments (red dots) and distribution of average total house price (blue dots) in Beijing (a) and Shanghai (b). 
}
\label{fig:house_price_freq}
\end{figure}

The distribution of house price in both cities 
manifest a power-law tail after the saturation-effect region (i.e., roughly after 4 million RMB in Beijing, and 3 million in Shanghai, see Fig. \ref{fig:house_price_freq}). The distribution of house price is similar to the distribution of wealth \cite{levy1997new,ding2007power} that follows a power-law, and this further supports that using house price is relatively reasonable to approximate income. 
In addition, the distribution of average house price and the distribution of total price of all apartments are in the same shape (see Fig. \ref{fig:house_price_freq}).  
This implies that the aggregation process to the location level can be a reasonable coarse-graining. 

\subsubsection{Detecting Home and Work Locations of Users}
A critical step of this study is to identify the home and work locations of dockless sharing bike users, as there is no ground truth in the original dataset. 
We use the following algorithm that is commonly applied in cellphone data analysis \cite{alexander2015origin,liu2022revealing,xu2017clearer,dong2016population} to make estimations. 
For each user, we count arrival locations of trips made between 22:00 to 5:00 of the next day, and departure locations of trips between 5:00 to 9:00. We then order locations descendingly by the appearance frequency of the user, and associate the location with the highest frequency as the home of the user, provided there is a ``Commercial House'' POI in this place; otherwise, check the next highest one and so on. 
We implement a similar algorithm to detect the working location (or more precisely, the main location of cycling activity during the day) for each user. We count departure and arrival locations of trips between 9:00 to 22:00 and associate the location with the highest appearance frequency as the work location of the user. The POI information is not applied for double checking, since the workplace can be of various POI types (e.g., restaurants, hotels, shopping malls). 

\begin{figure}[!t] \centering
\subfloat[]{\includegraphics[width=0.5\linewidth]{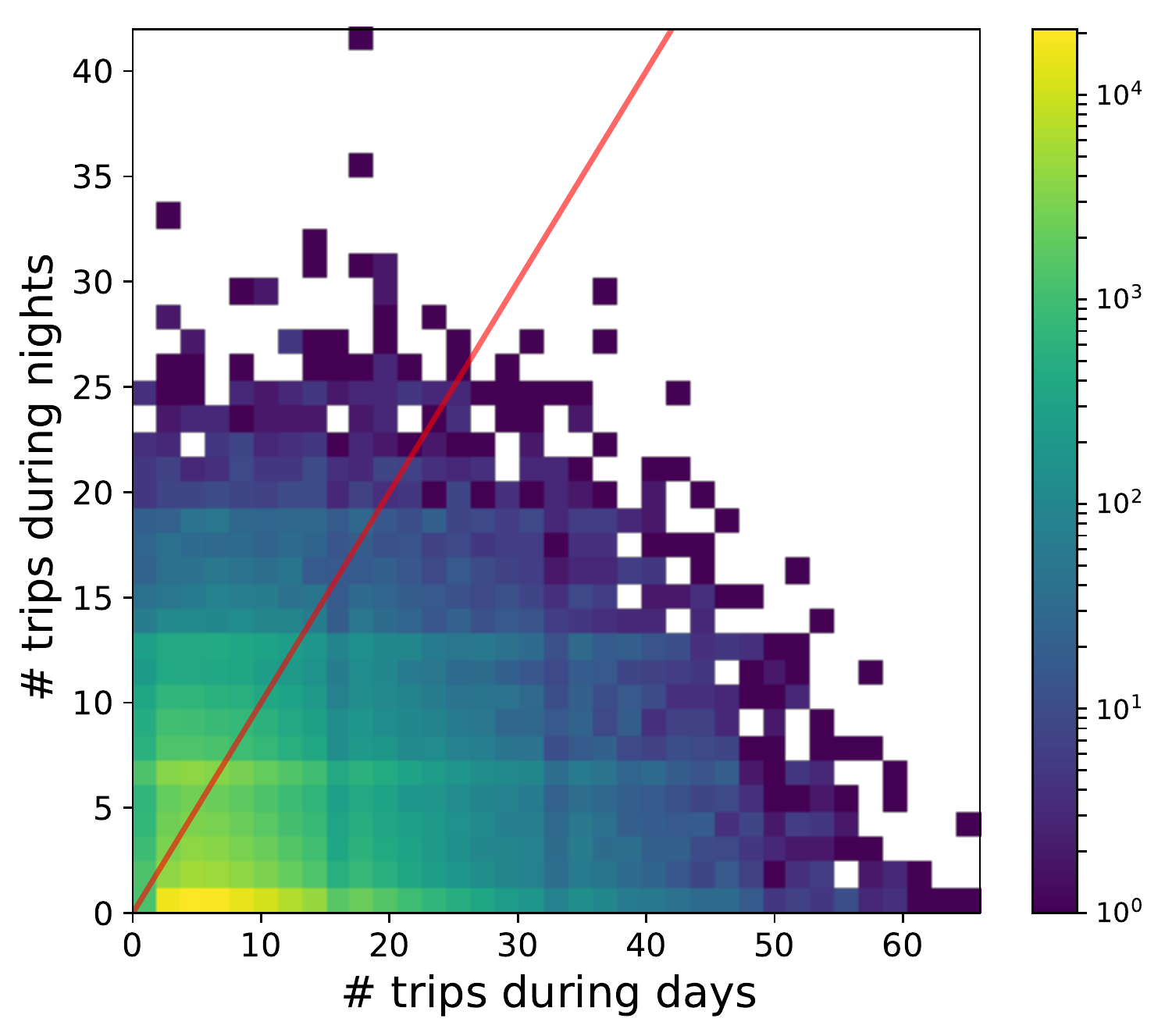}%
\label{fig:time_dist_beijing}}
\hfil
\subfloat[]{\includegraphics[width=0.5\linewidth]{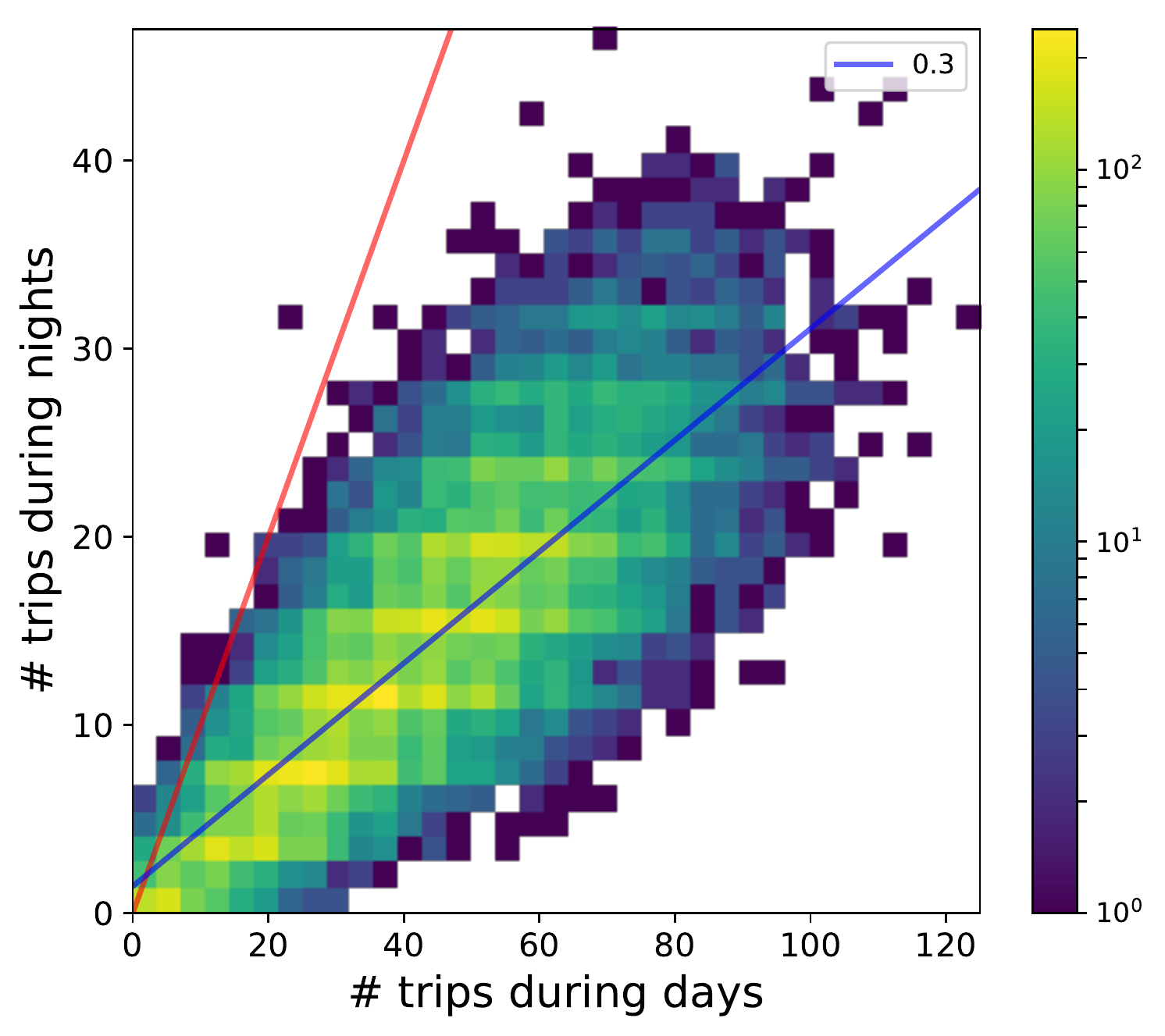}%
\label{fig:time_dist_shanghai}}
\caption{The frequency distribution of biking trips during daytime and nighttime of all user in Beijing (a) and Shanghai (b). The red line is the $y=x$. 
}
\label{fig:time_dist}
\end{figure}

With such a time division, we also divide the trips into daytime and nighttime. 
Users in Shanghai tend to have a linear relationship between the daytime and nighttime travels with a coefficient of $0.3$. While, such a linear trend is absent in Beijing and there are some extreme day users or night users with either more trips during nights than days or the vice versa, which is non-existing in Shanghai (see Fig. \ref{fig:time_dist}). There are some extreme day users or night users, though recent advance indicate positive impact of the length of daylight on cycling activities \cite{wessel2022cycling}, this suggest that urban characteristics might play a stronger role on cycling profiles. 

\begin{figure}[b] \centering
\subfloat[]{\includegraphics[width=0.42\linewidth]{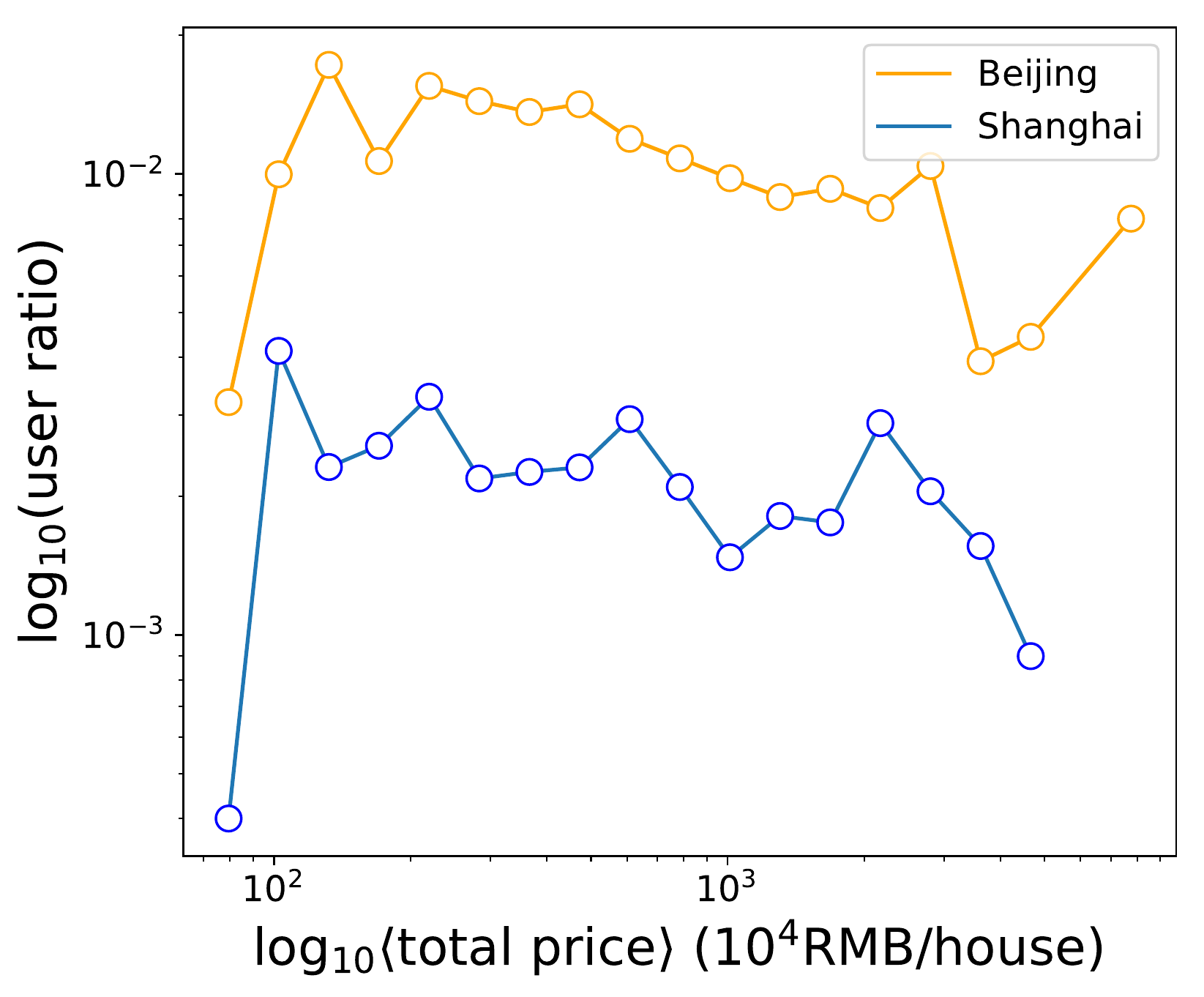}%
\label{fig:ratio_total_beijing}}
\hfil
\subfloat[]{\includegraphics[width=0.5\linewidth]{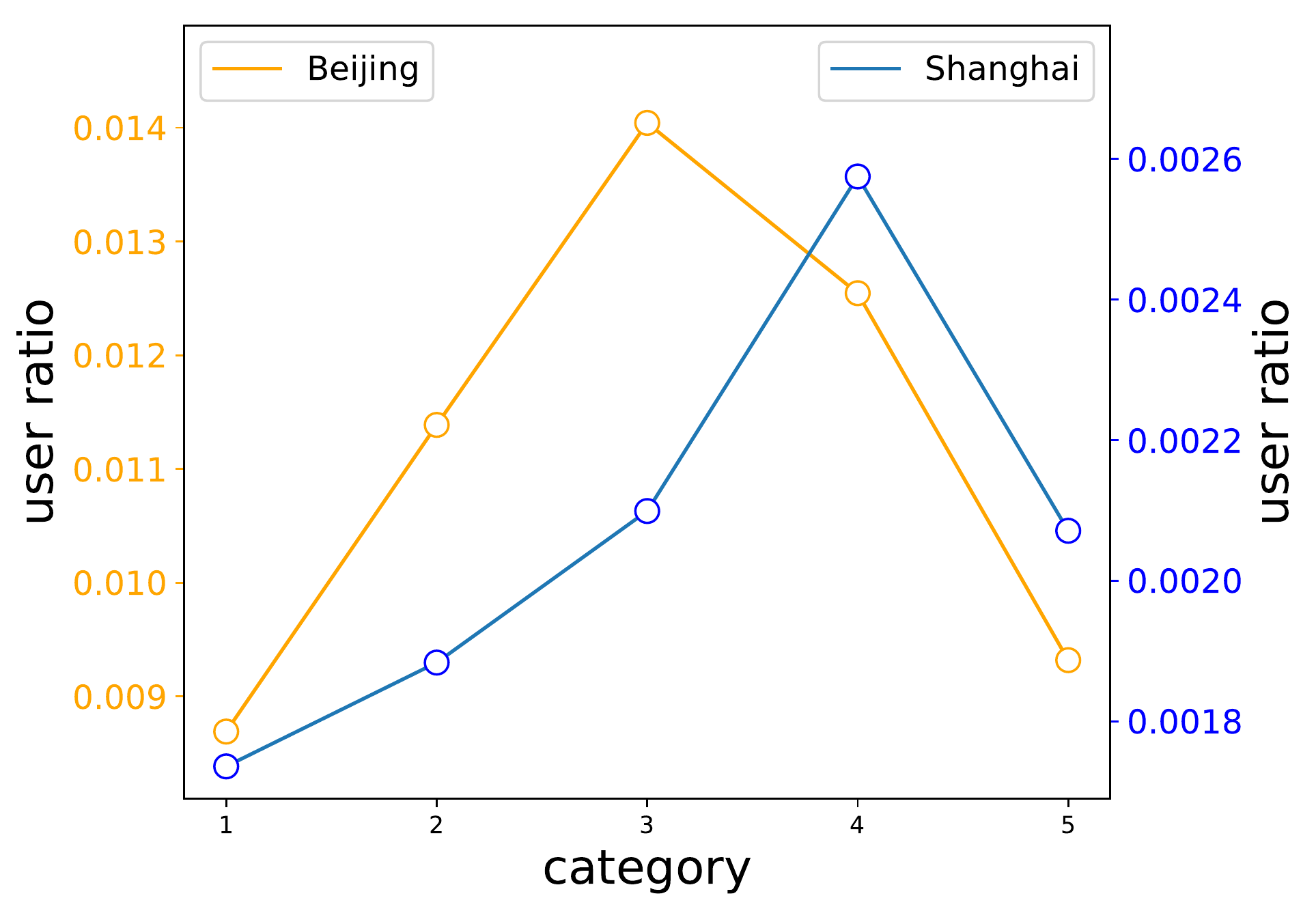}%
\label{fig:ratio_total_shanghai}}
\caption{The relation between user ratio $r_i$ and total house price over locations (a) and in each income category (b). The 1$^{st}$ and 5$^{th}$ category corresponds to richest and poorest neighborhoods, respectively. \label{fig:ratio_total_price}}
\end{figure}

Based on above detection algorithms, we can obtain the number of users living in each location. As a supplement, the residential population of locations in Beijing and Shanghai is obtained from the WorldPop (\url{www.worldpop.org}). 
Then we can calculate the ratio of the number of users to the residential population in each location $r_i$. 
When associating $r_i$ and average total house price of locations, we can observe that locations with a relatively large $r_i$ have a moderate house price, 
and there is a slight declining trend with fluctuations when house price increase (see Fig. \ref{fig:ratio_total_price}a). 
This indicates that people in places with a higher house price have a weaker tendency to use dockless sharing bikes. When we look at the user ratio in different income categories, it is obvious that the 1$^{st}$ category that are of the highest house price is of the lowest user ratio, and the user ratio goes in a parabolic shape towards lower-income categories (see Fig. \ref{fig:ratio_total_price}b). For example, the lowest income category (i.e., the 5$^{th}$ one) in Beijing is also of a pretty low user ratio. 

\subsubsection{Identification of Socioeconomic Status of Users}
Based on the Loubar method introduced in Section \ref{sec:lorenz_curve}, we divide all locations into different income categories based on their average house price. The range of house price, the number of locations (\#) and its proportion (\%) in each category are shown in Table \ref{tbl:income_level} for Beijing and Shanghai. The division process of the Loubar method is presented in Fig. \ref{fig:Lorenz_Curve} for Beijing and Shanghai, where the four dividing tangential lines for each city start from the reddest line on the right to those lighter red lines on the left. Then we associate users to the income category with the level identified for the location where they reside. 

\begin{figure}[!t] \centering
\includegraphics[width=0.6\linewidth]{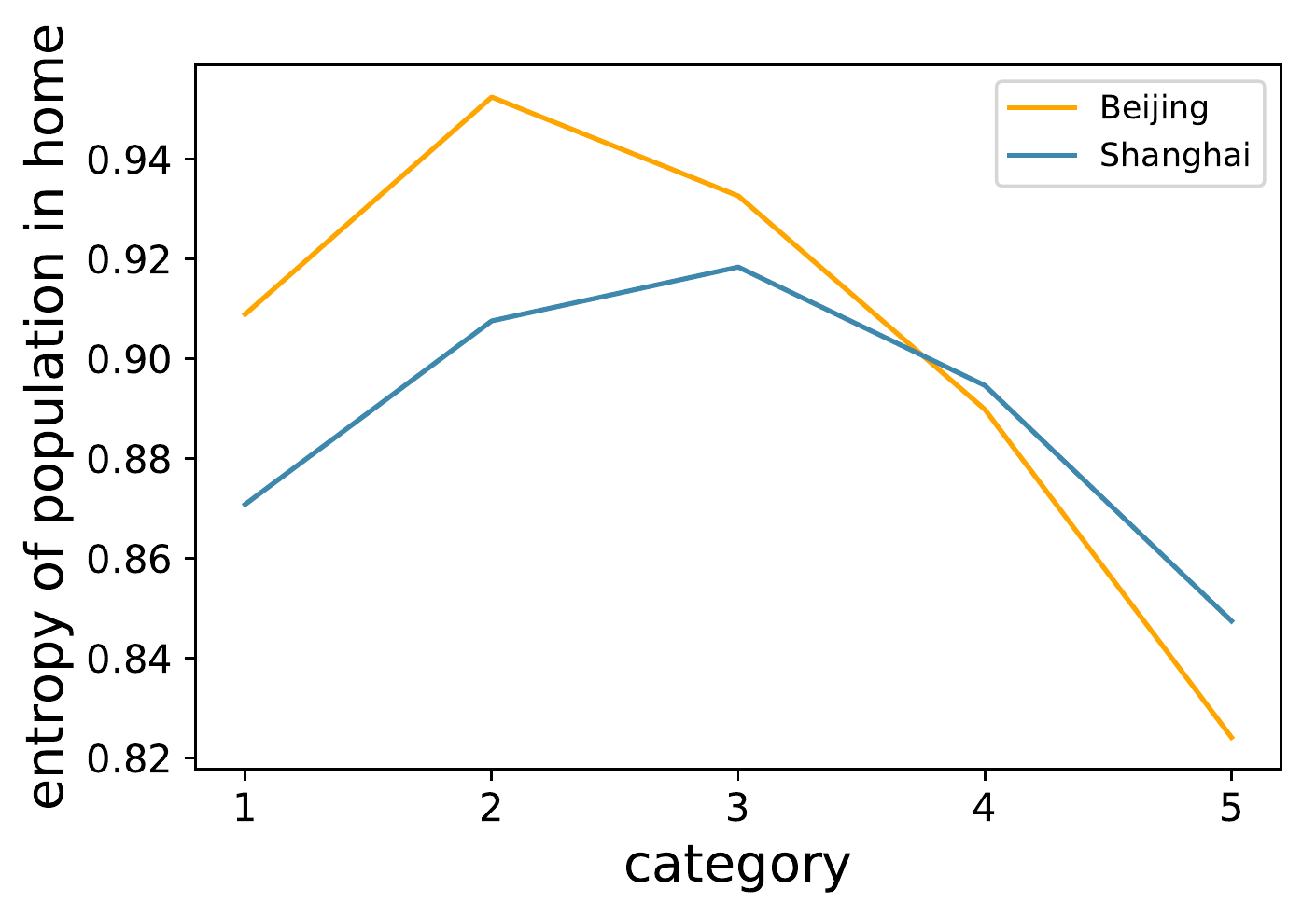}
\caption{The residence diversity of users in different income categories in Beijing and Shanghai. 
\label{fig:entropy_population}}
\end{figure}

Another interesting question is that for users in each income category, are they concentrating in fewer locations or living more evenly over locations. This can be quantified by the residence diversity as introduced in Section \ref{sec:entropy}. We observe a bell shape in both cities with richest and poorest people all living relatively concentrated in fewer locations, and the poorest people are the most concentrated group with the lowest residence diversity (see Fig. \ref{fig:entropy_population}). 

\section{Results\label{sec:results}}
In this section, we explore mobility patterns at both the individual and collective level for users in different income categories. The methods and indicators mentioned in Section \ref{sec:approach} are used for the analysis.

\subsection{Individual Mobility Patterns}

\begin{figure}[b]\centering
\subfloat[]{\includegraphics[width=0.5\linewidth]{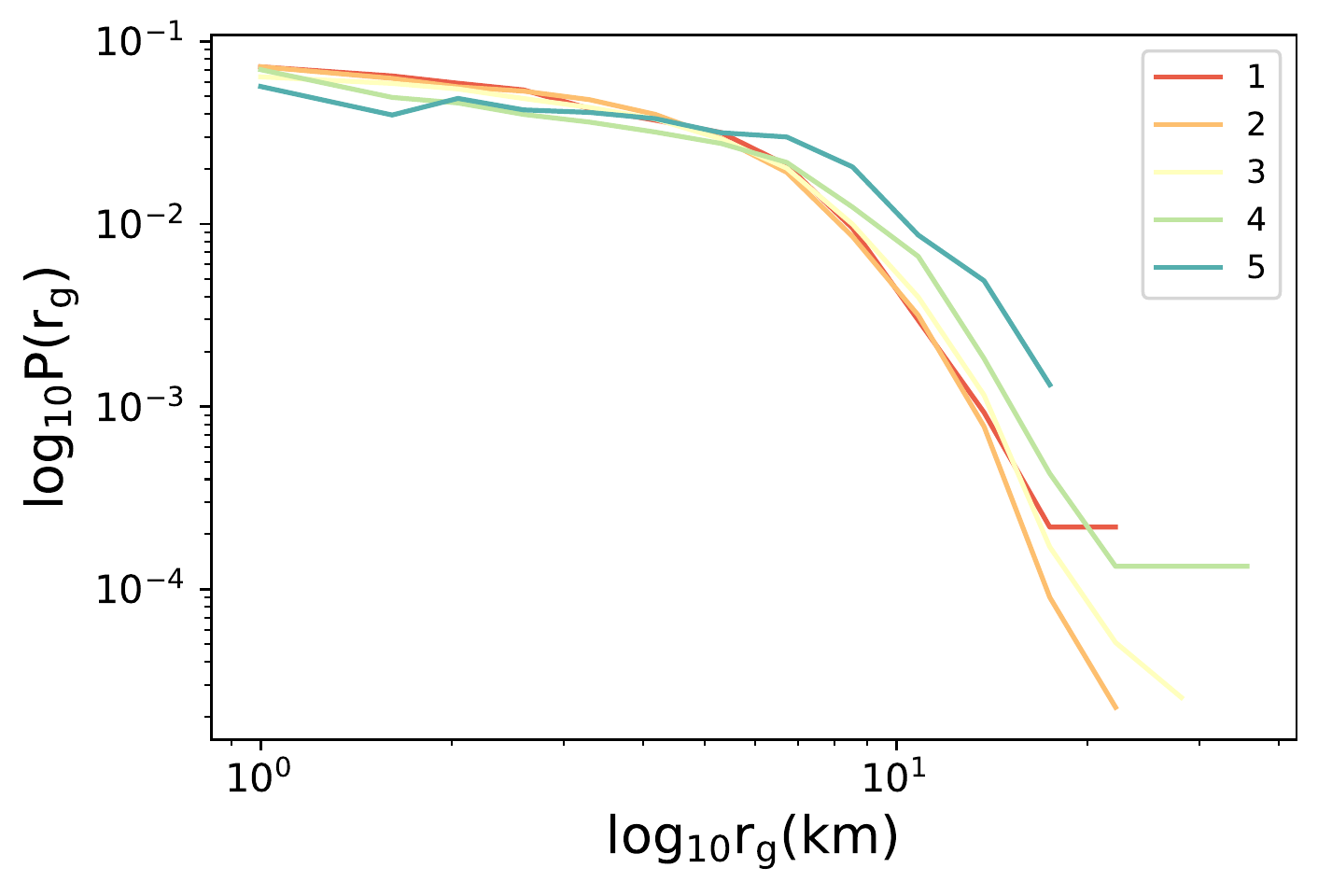}%
\label{fig:radius_of_gyration_beijing}}
\hfil
\subfloat[]{\includegraphics[width=0.5\linewidth]{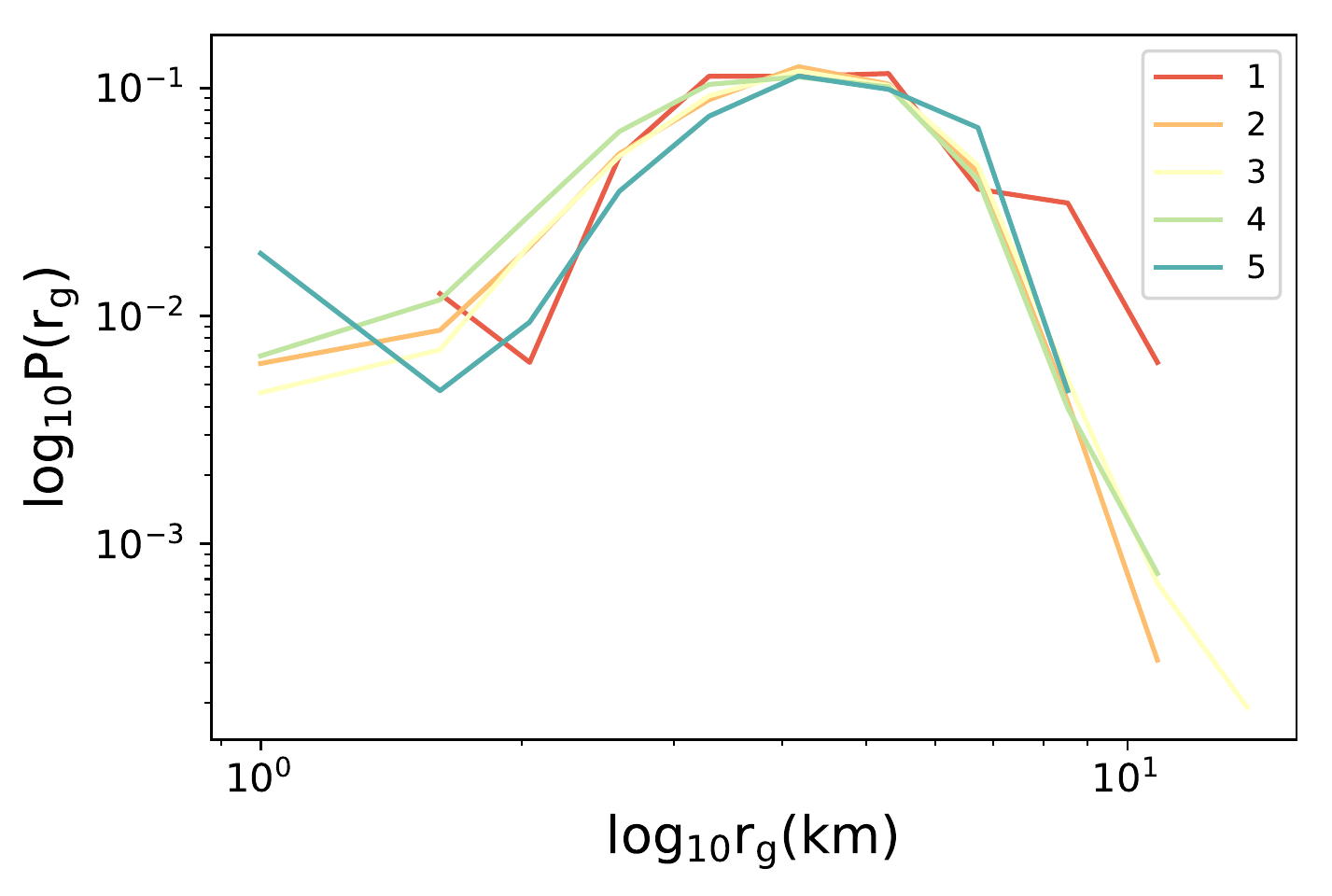}%
\label{fig:radius_of_gyration_shanghai}}
\caption{The radius of gyration for users in different income categories in Beijing (a) and Shanghai (b).}
\label{fig:radius_of_gyration}
\end{figure}

To study the individual mobility patterns of users different income categories, we first calculate the radius of gyration as described in Section \ref{sec:radius_of_gyration}. The radius of the gyration can reveal the average range of visited locations of a user to its center point. 
For both Beijing and Shanghai, we observe similar patterns for users in all five income groups (see Fig. \ref{fig:radius_of_gyration}). This indicates that humans are all following similar physical laws. However, the pattern is different between Beijing and Shanghai. In Beijing, a significant amount of people has a small radius of gyration, which suggests that they are using dockless sharing bikes only in relatively limited regions, while in Shanghai, there are more users with a larger radius of gyration, which suggests that they might ride bikes to explore more places distant from each other. The difference between cities are again would be influenced by various urban characteristics. 

\begin{figure}[!t]\centering
\subfloat[]{\includegraphics[width=0.5\linewidth]{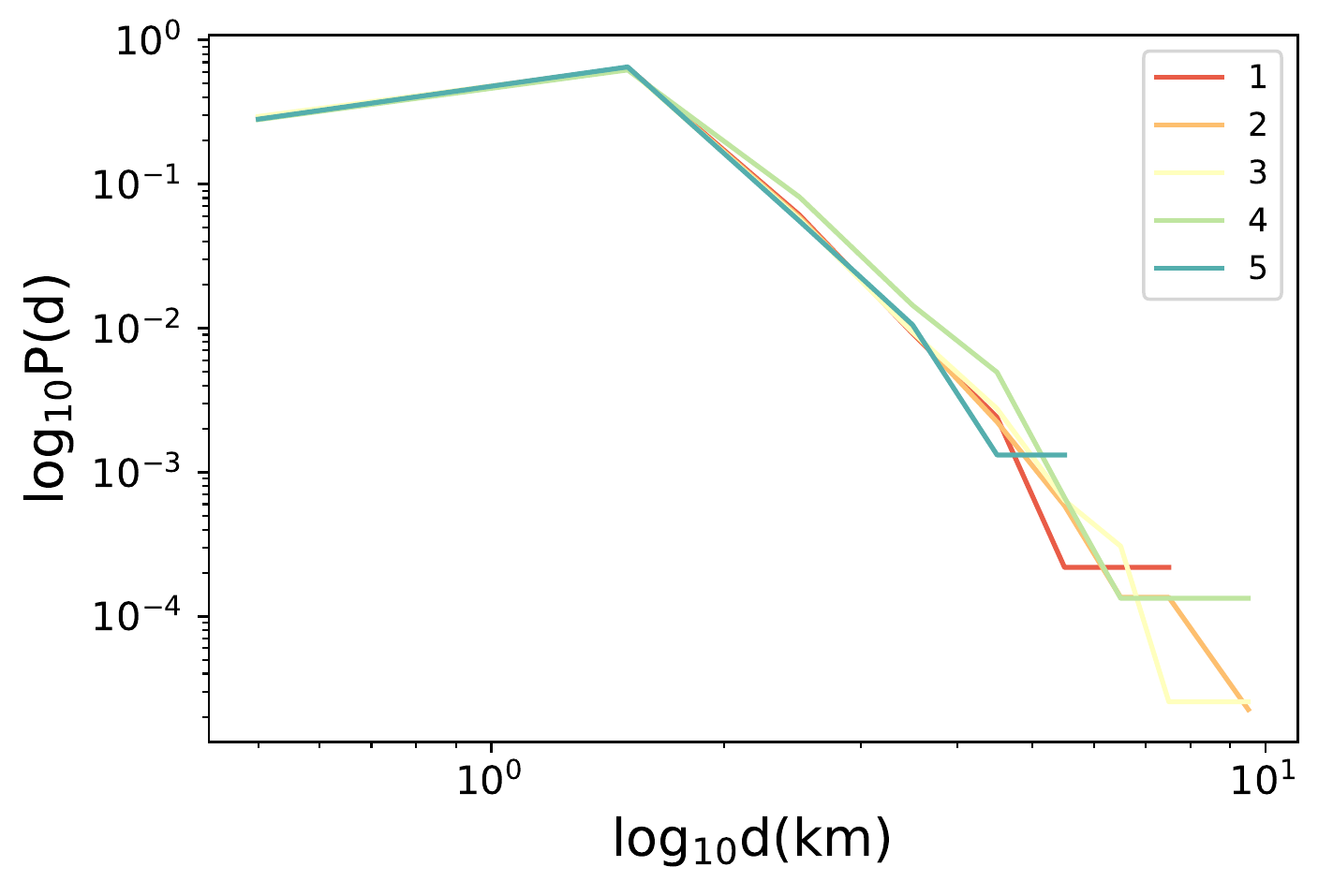}%
\label{fig:travel_distance_beijing}}
\hfil
\subfloat[]{\includegraphics[width=0.5\linewidth]{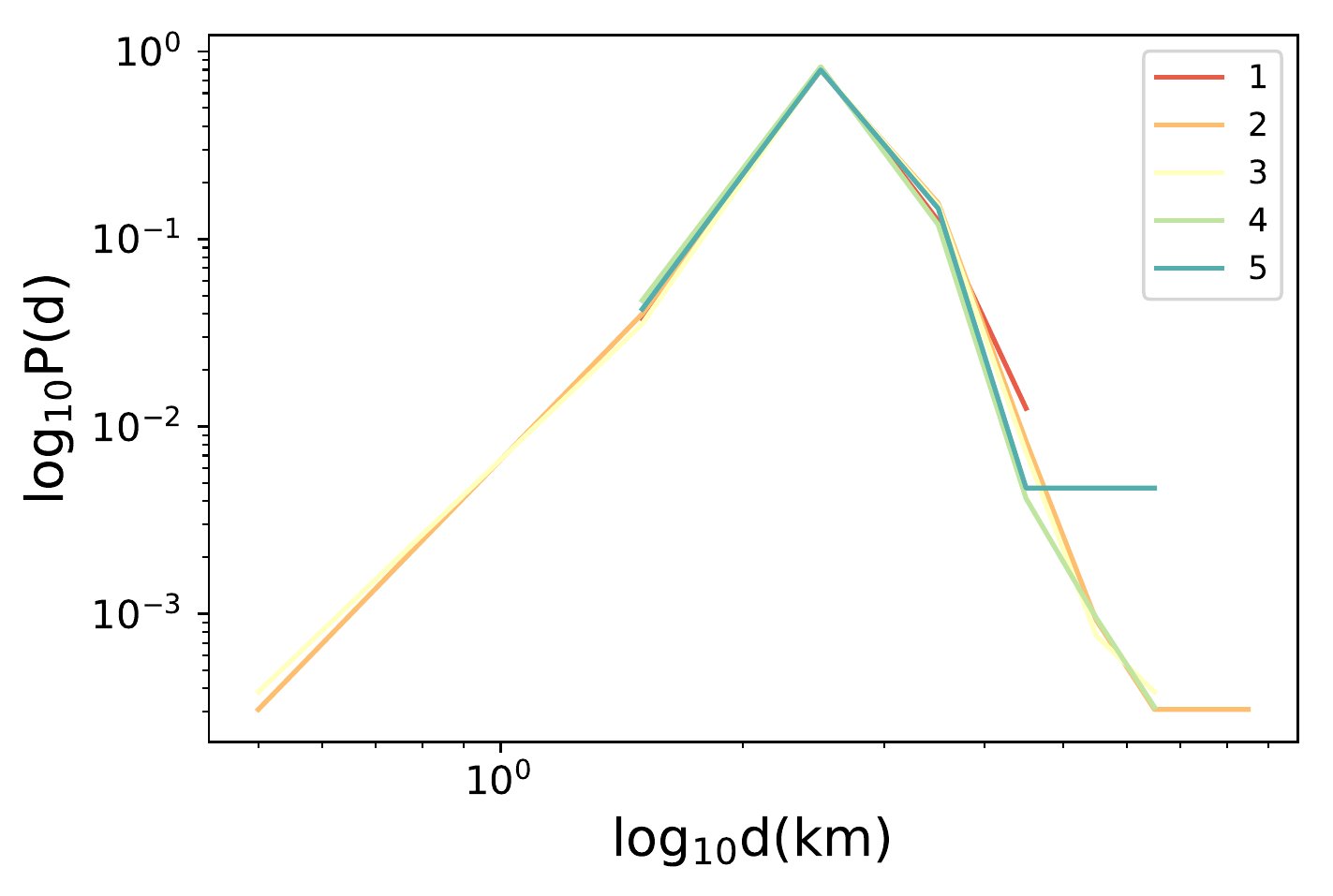}%
\label{fig:travel_distance_shanghai}}
\caption{The average travel distance of users in different income categories in Beijing (a) and Shanghai (b).}
\label{fig:travel_distance}
\end{figure}

Similar phenomenon is observed on the distribution of average travel distance in Beijing and Shanghai (see Fig. \ref{fig:travel_distance}), where the difference between users in different income categories are not very huge. 
As cycling is a physical exertion, thus we expected that the average travel distance by bikes should be closer to a normal distribution, which is indeed the case in Shanghai (see Fig. \ref{fig:travel_distance_shanghai}) but, surprisingly, not the case in Beijing (see Fig. \ref{fig:travel_distance_beijing}). The distribution of average travel distance exhibits a power-law tails which indicates that there is a non-negligible fraction of users on average ride for several up to roughly ten kilo meters when they user dockless sharing bikes. Previous studies suggest that dockless sharing bikes are mainly used as short-distance commuting connectors \cite{chen2019analyzing}, but such a discovery in Fig. \ref{fig:travel_distance_beijing} suggest that many users in Beijing might strongly rely on dockless sharing bikes for long-distance trips that goes beyond commuting connection.  

\begin{figure*}[!htbp] \centering
\includegraphics[width=\linewidth]{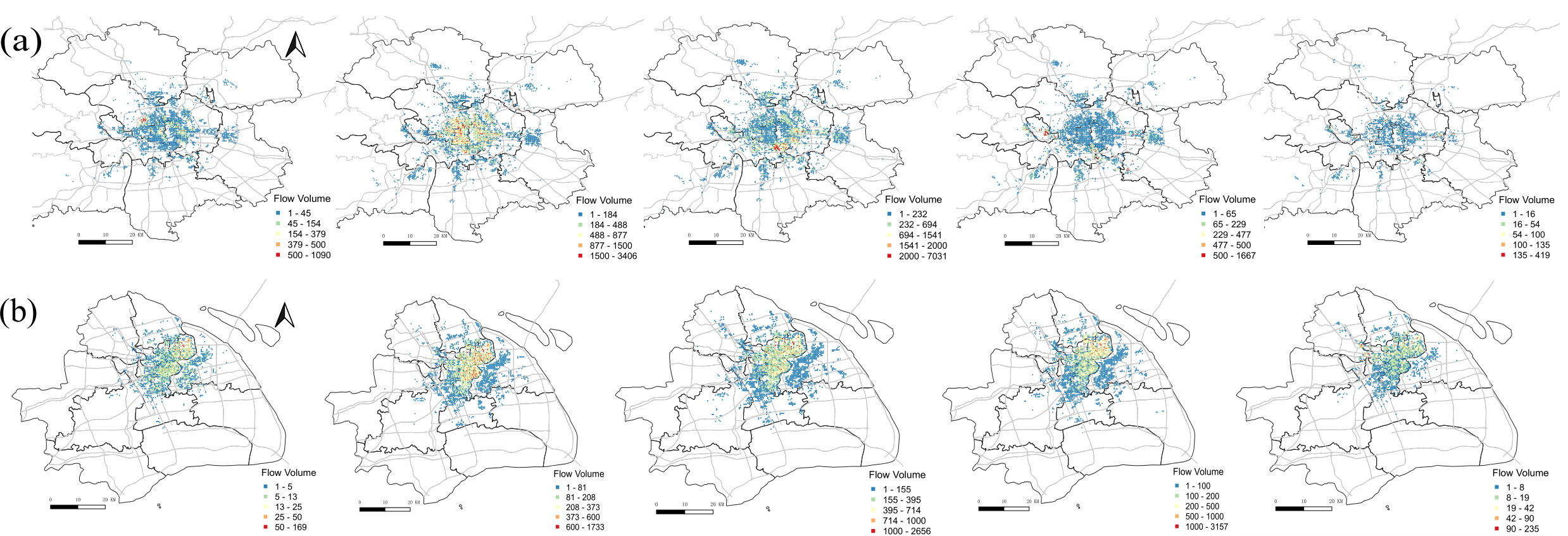}
\caption{The spatial distribution of flow volume in Beijing (a) and Shanghai (b) for different income categories. 
}
\label{fig:spatial_flow_income}
\end{figure*}

\subsection{Collective Mobility Patterns}

\begin{figure}[!tb] \centering
\includegraphics[width=2.5in]{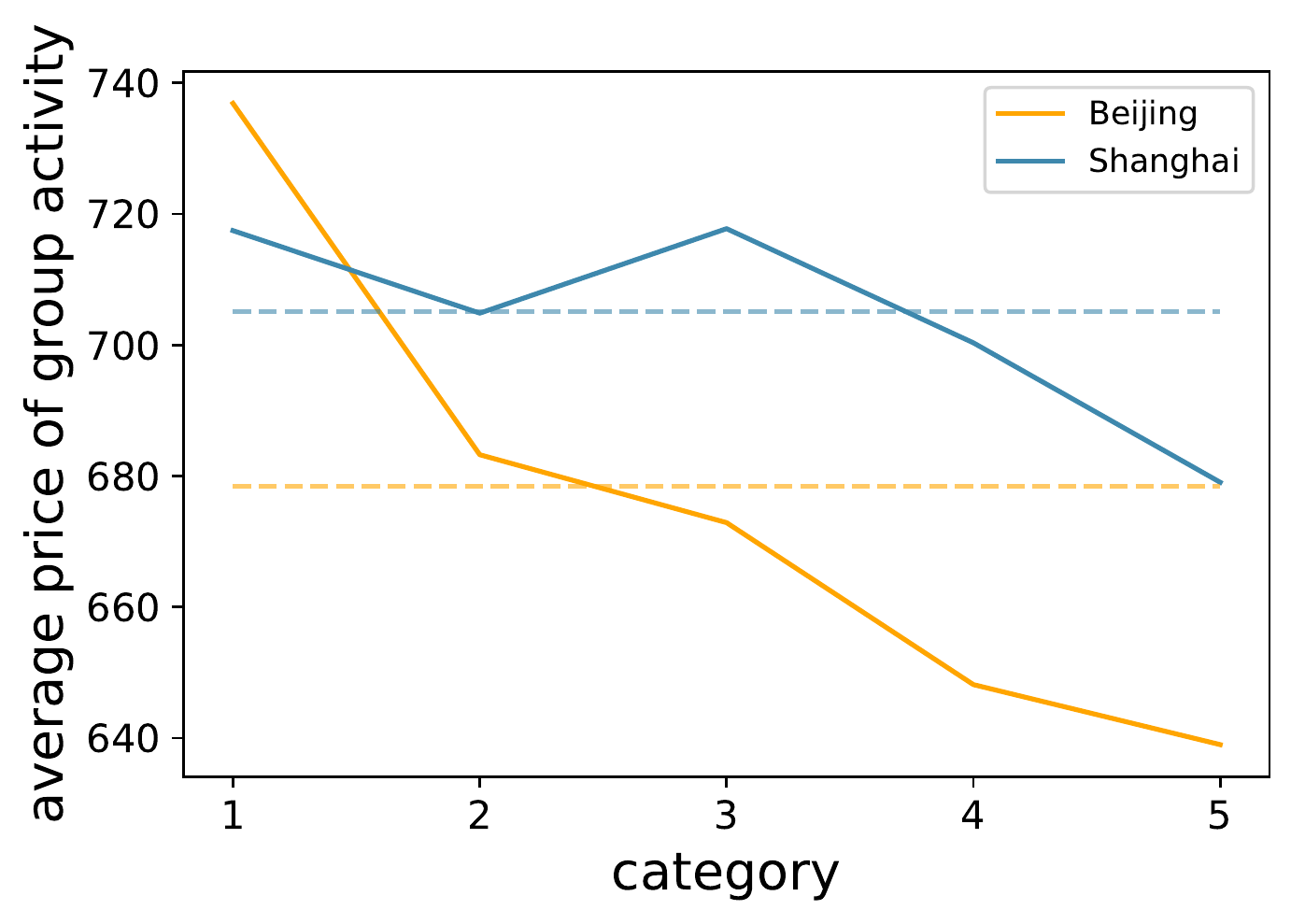}
\caption{The average house price of locations visited by users from different income categories in Beijing and Shanghai. The dashed lines present the average house price of visited locations by all users in the whole city, which is served as a baseline. \label{fig:price_group_activity_income}}
\end{figure}

We find that the spatial distribution of flow volume by all users in different income categories are dissimilar and heterogeneous in Beijing (see Fig. \ref{fig:spatial_flow_income}), and hotspot locations that attract large flow volume located in different districts for different groups. This indicate that users with different income tend to visit different locations in Beijing. While in Shanghai, the spatial distributions are relatively similar and hotspot locations are not too far away over different income categories (see Fig. \ref{fig:spatial_flow_income}). 
We further calculate the average house prices of locations visited by users from different income categories in Beijing and Shanghai (see Fig. \ref{fig:price_group_activity_income}). 
We observe a stronger decrease trend in Beijing than in Shanghai, which suggest that users in lower-income groups tend to visit less flourishing places. This might be an indication for the existence of potential income segregation. 

\begin{figure}[!t] \centering
\includegraphics[width=2.5in]{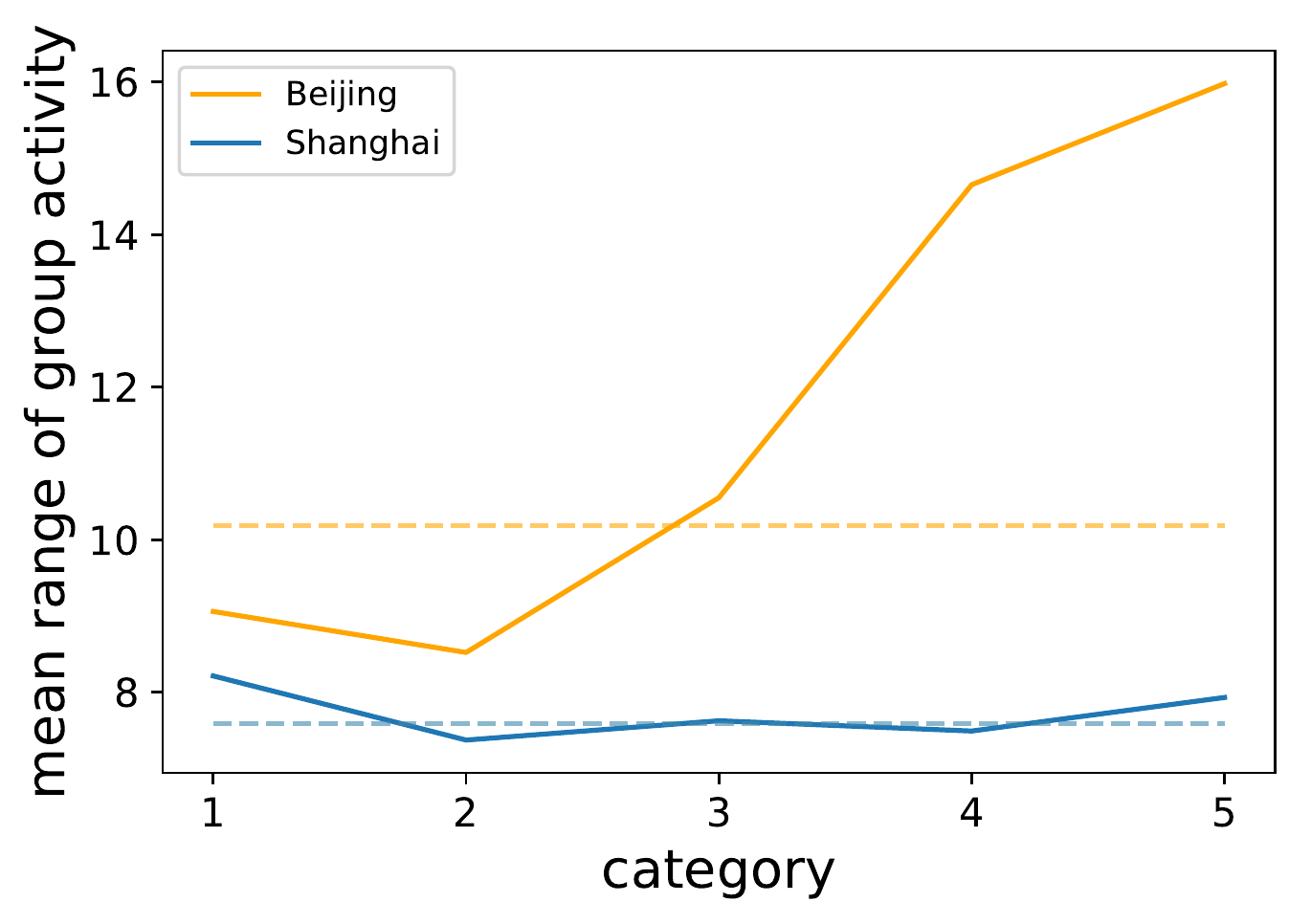}
\caption{The average range of locations visited by users from different income categories in Beijing and Shanghai. The dashed lines present the average value of all locations that have been ever visited, which is served as a baseline.  \label{fig:group_activity_range_income_level}}
\end{figure}

By applying Eq. (1)  to all locations visited by users in each income category, we can quantify the collective dispersion degree from the mass center $r_c$. We find that mass centers of different groups are all close to the city center in both Beijing and Shanghai, but the average range are increasing rapidly in Beijing (see Fig. \ref{fig:group_activity_range_income_level}). This is due to the fact that, in Beijing, the lower-income groups have fewer visitations to locations near the center and more visitations to locations at the fringe of the city (see the fifth sub-figure of Fig. \ref{fig:spatial_flow_income}a), 
thus its average range is larger. 
In comparison, the spatial distribution of locations visited by users from different categories in Shanghai are relatively similar (Fig. \ref{fig:group_activity_range_income_level}). 

\begin{figure}[!b] \centering
\includegraphics[width=2.5in]{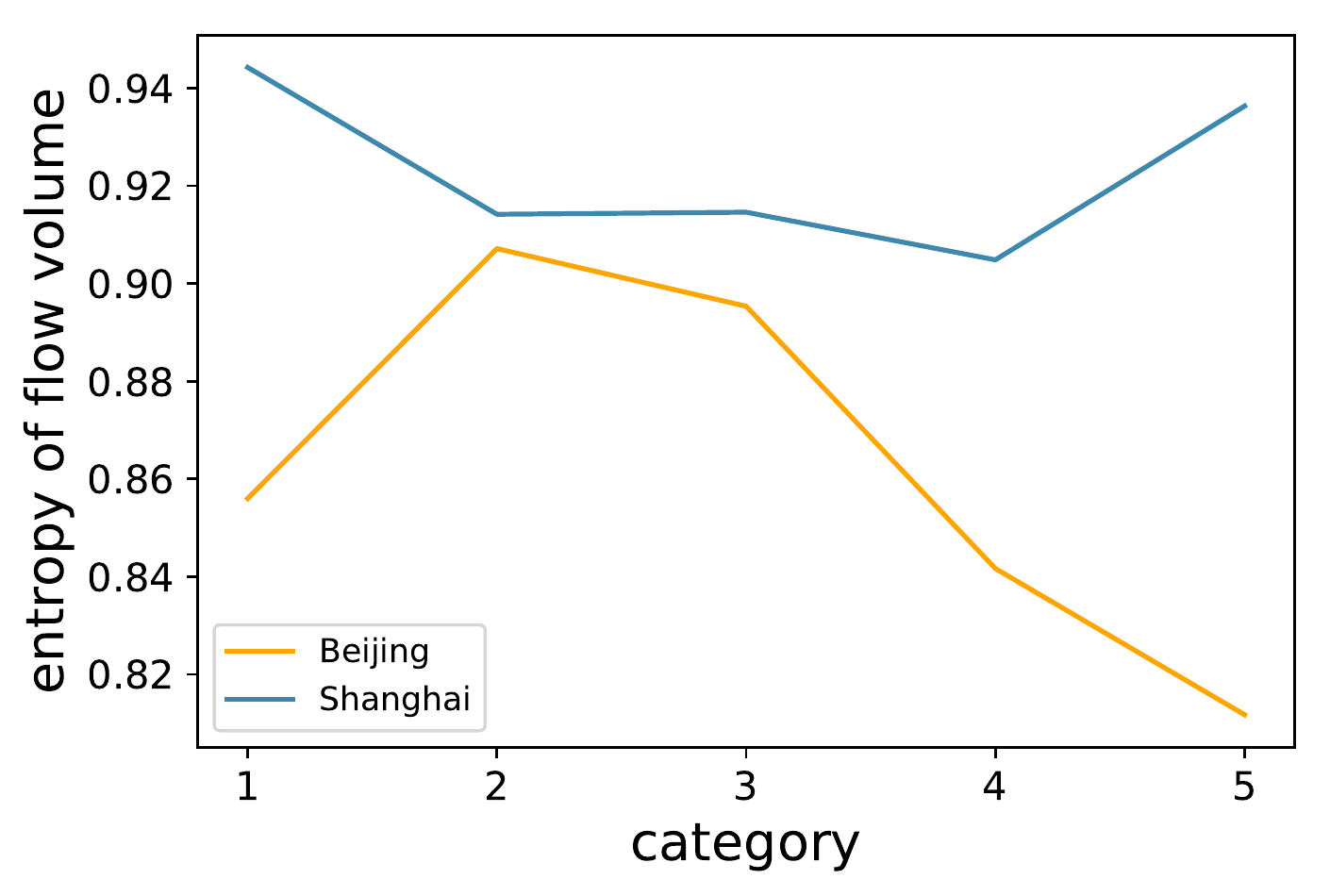}
\caption{The entropy of flows for different income levels in Beijing and Shanghai. \label{fig:entropy_flow}}
\end{figure}

To better quantify the flow diversity of each category, 
we apply Eq. (2) introduced in Section \ref{sec:entropy}. The diversity would be lower if flows are more concentrated in a few locations, and would be higher if distributed more evenly over all locations. 
In Beijing, we observe a bell shape with both the highest and lower income categories exhibiting a more concentrated visitation patterns indicated by a lower diversity value (see Fig. \ref{fig:entropy_flow}). This implies that certain locations are quite attractive to users in these groups. 
While, an opposite ``U'' shape is observed in Shanghai, where both the highest and lowest income categories have higher diversity. And the flow diversity of different categories in Shanghai are all larger than in Beijing. For example, when comparing the fifth sub-figure of both Fig. \ref{fig:spatial_flow_income}a and Fig. \ref{fig:spatial_flow_income}b, we can find that although the largest volume in Beijing is larger than in Shanghai, but the volume is quite concentrated in a few locations in Beijing, while in Shanghai, the flow are evenly distributed over many locations. 

\begin{figure*}[!t]\centering
\includegraphics[width=\linewidth]{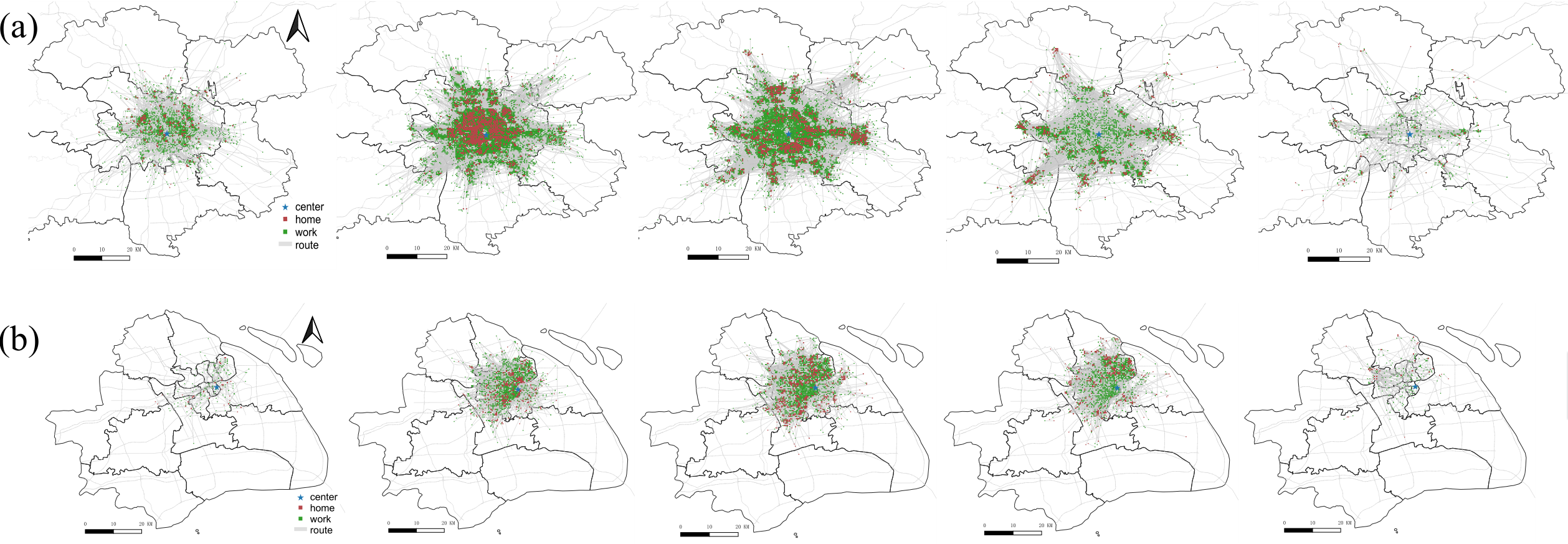}
\caption{The spatial distribution of inferred home and work locations of users in different income categories (1$^{st}$ to 5$^{th}$ from left to right) in Beijing (a) and Shanghai (b).}
\label{fig:spatial_home_work}
\end{figure*}

We further explore the spatial distribution of home and work locations of users in different income categories for Beijing and Shanghai (see Fig. \ref{fig:spatial_home_work}). Each pair of connected red-green squares indicate the home and work locations of a user. When a place is both a home location for some users and a work location for other users, we still mark it as red. 
We observe that there are more work locations than home locations in all five groups, and work locations are more dispersed. When comparing between different groups, the spatial distribution of home and works locations are quite dissimilar. But we can still observe a tendency of working near the city center for 
all groups in both cities (see Fig. \ref{fig:spatial_home_work}). 

\begin{figure}[!t] \centering
\includegraphics[width=0.55\linewidth]{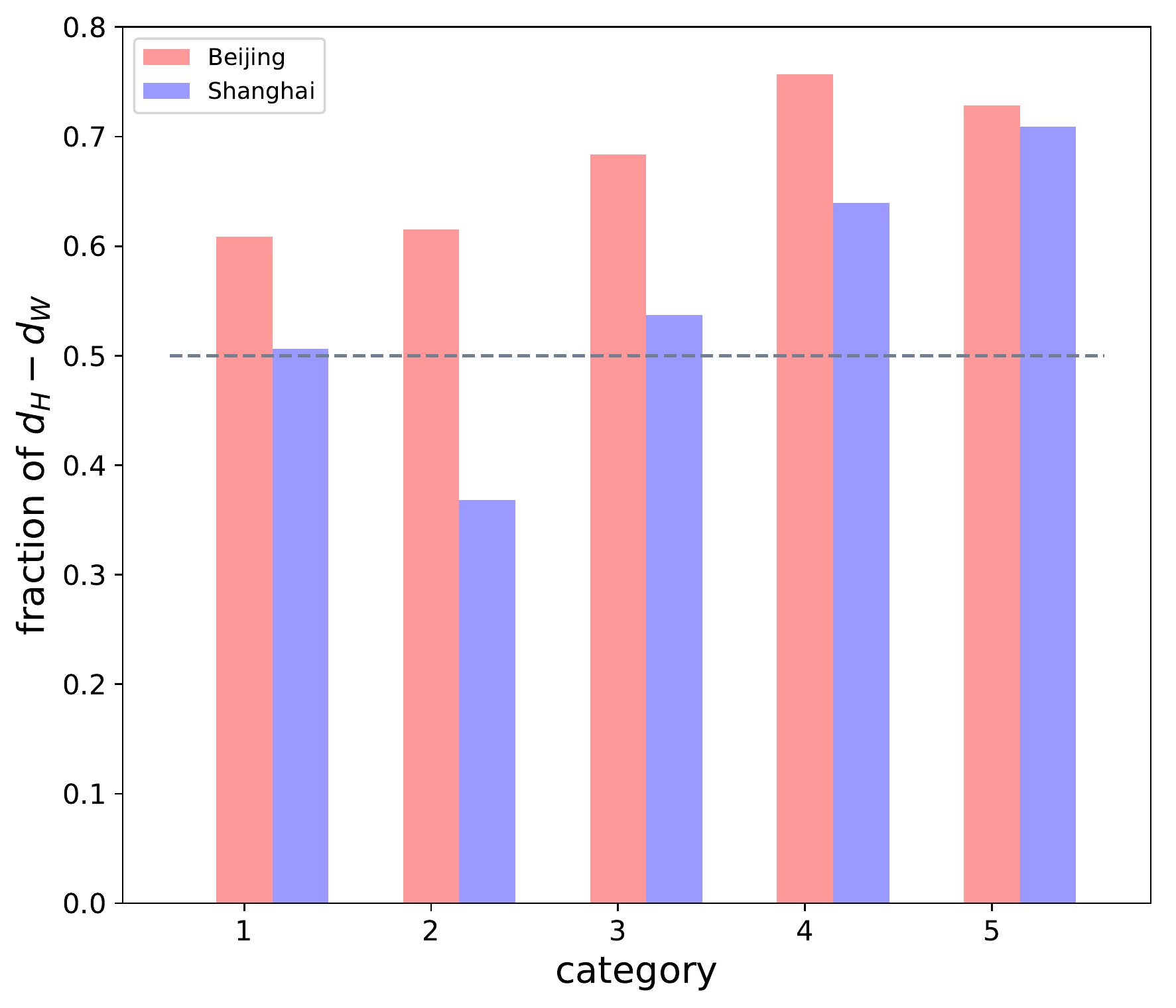}
\caption{The proportions of commuting direction of users from different income categories in Beijing and Shanghai. \label{fig:work_trip_outward}}
\end{figure}

The direction of the commuting trip from home to work location can be a reflection of the extend of monocentric versus polycentric \cite{remi2013emergence}. 
We calculate the distance from the home location to the city center (denoted as $d_H$) and from the work location to the city center ($d_W$), we can observe that most people in each category (except the 2$^{nd}$ category in Shanghai) are commuting towards the direction to the city center (though their work locations might not really near the city center), which is indicated by a larger fraction of positive value of $d_H-d_W$ (see Fig. \ref{fig:work_trip_outward}). In addition, lower income groups roughly have a higher fraction of having a work location closer toward the city center. This might implies that most people might still live in places with a house price lower than their work locations, as the spatial gradient of unit house price is quite obvious (see Fig. \ref{fig:spatial_price}) and thus a farther away location is usually of a lower price. 

\section{Discussion \label{sec:conclusion}}
In this work, we study the relation between various mobility patterns of and the income level of dockless sharing-bike users in Beijing and Shanghai. 
We find that in terms of the individual mobility patterns, users in all five income categories have a similar behavior in each city but may behave differently across cities due to influence of various urban characters, which is demonstrated by the distribution of radius of gyration and the average travel distance. While, at the collective level, most mobility patterns are quite different over income categories. 
The most visited locations (i.e., hotspots) of each group are quite different from each other in Beijing, and relatively similar in Shanghai. But the trend that users from lower-income categories of visiting less flourishing places are all present in both cities with a different extend. More quantitatively, the average range of cycling activities of lower-income groups in Beijing are larger which suggest that they have fewer visitations to locations near the central regions. In comparison, the average range of cycling activities of users from different income categories in Shanghai are more similar. 
We also observe that cycling activities of the highest and lower-income groups in Beijing are more concentrated as indicated by a lower diversity, while in Shanghai, an opposite pattern is observed and the diversity is larger than in Beijing. This might imply different usage dynamics. 
In addition, after detecting home and work locations of users in different income categories, we further reveal that there is increasing trend of commuting towards the city center from higher high-income group to lower-income groups. 
We also discover that, during the period of the dataset, the user ratio is quite low for both higher- and lower-income groups (see Fig. \ref{fig:ratio_total_price}) and their residence diversity is also low (i.e., most users in these groups are concentrated in a few locations, see Fig. \ref{fig:entropy_population}), thus there is a great market potential in the neighborhoods and the need of designing better promotion strategies to related groups and locations, including local advertising, giving special offers, or free trials. 

Above discoveries also can be useful to assist better design of company promotion strategies and related governmental policies on green transportation. For example, based on mobility patterns, the cycling network can be better designed; and the government give more attentions to lower-income groups. Quite recently, electric sharing scooters \cite{cao2021scooter} and electric dockless sharing bikes are more and more popular, 
which are usually manually deployed by the company after recharging. Based on house price, ratio of dockless sharing bike users, and their commuting patterns, a prediction for potential users of electric dockless sharing bikes is viable. 

In this study, there are some limitations posed by several factors. Due to limitation of data collection, currently the home and work locations are estimated by from travel behaviors but with no ground truth to test with, thus some mis-identifications are inevitable. And the 
period of the dataset is relatively short, when there are more data for a longer period available, seasonal fluctuations of cycling behaviors and related mobility patterns can be better analysed. 
In addition, 
we discover some fundamental differences present between Beijing and Shanghai, with data of more cities available in the future, a more comprehensive typological study \cite{Barbosa2021hundred,louf2014typology} can be conducted with a similar framework here.

\section{Methods\label{sec:approach}}

\subsection{Loubar algorithm for objective classifications\label{sec:lorenz_curve}} 
\begin{figure}[!htbp] \centering
\subfloat[]{\includegraphics[width=0.5\linewidth]{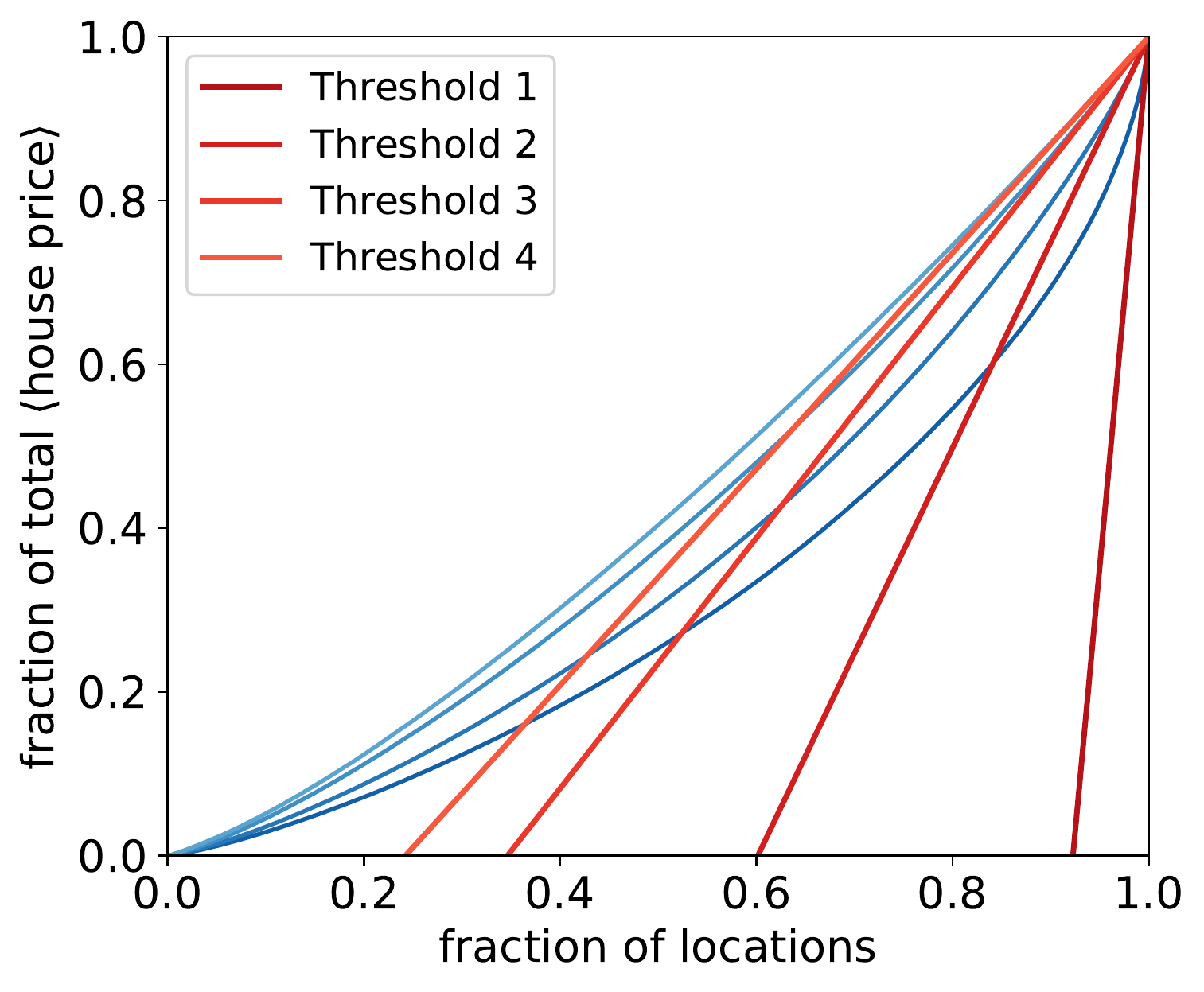}%
\label{fig:lorenz_curve_beijing}}
\hfil
\subfloat[]{\includegraphics[width=0.5\linewidth]{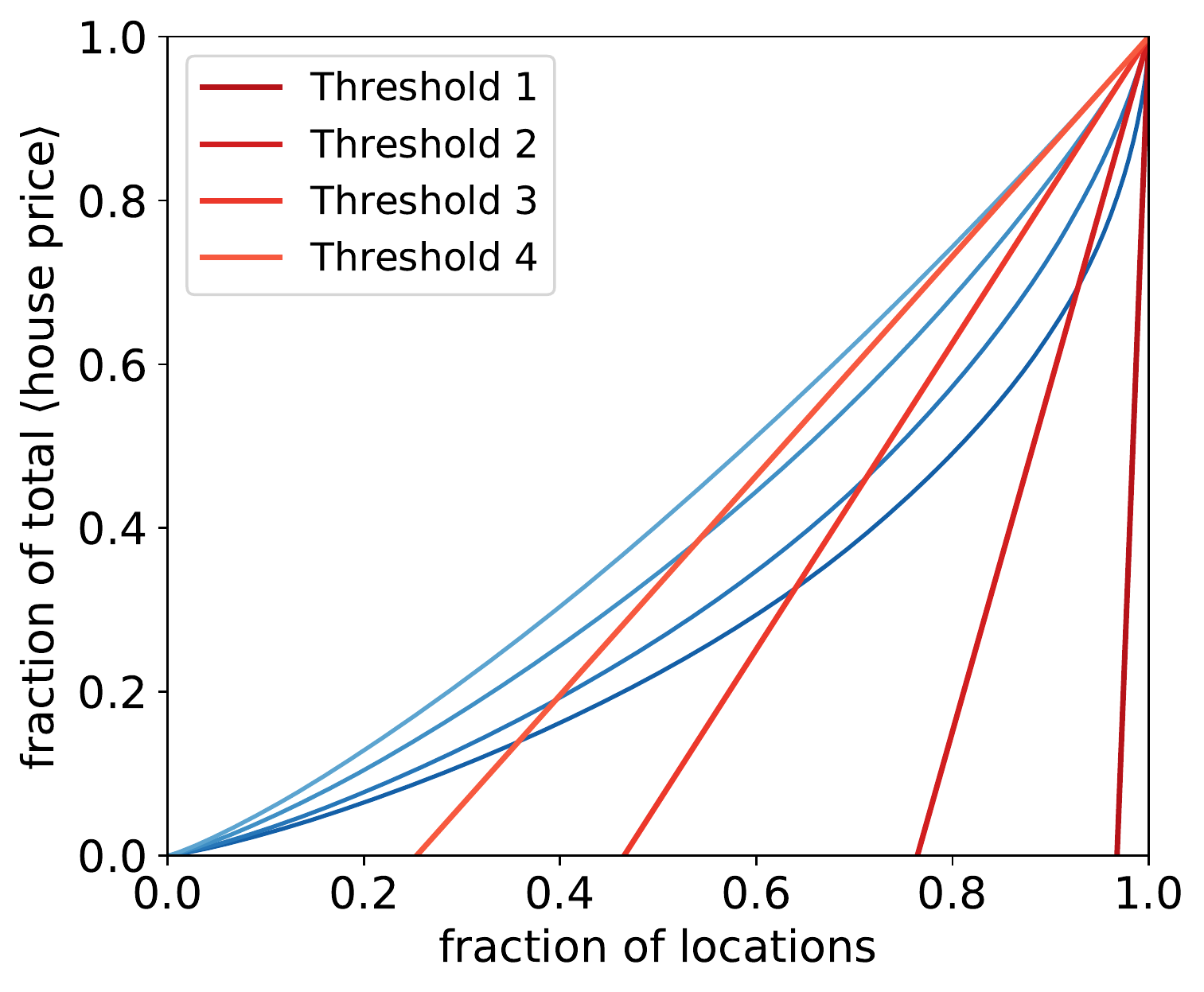}%
\label{fig:lorenz_curve_shanghai}}
\caption{The classification process of the Loubar method based on average house price of locations in Beijing (a) and Shanghai (b).}
\label{fig:Lorenz_Curve}
\end{figure}

To objectively identify the income level of users, which is approximated by average house price of locations where they reside, we use an iterative nonparametric Loubar method based on the derivative of Lorenz curves \cite{bassolas2019hierarchical,louail2014mobile}. 
The algorithm first sort all locations ascendingly based on their average house price, and normalizes them by the total number of locations. 
In other words, the value $n$ on the x-axis represents the first $n$ percent of locations with lowest average house price (see Fig. \ref{fig:Lorenz_Curve}).  The normalized cumulative value of house price is the y-axis. 
We then draw the tangential line of the cumulative curve at point $(1,1)$, and divide all locations by the intercept point of the tangential line on the x-axis (see reddest line on the far rightest in Fig. \ref{fig:Lorenz_Curve}). Locations with a larger value than the intercept point are assigned to first income category, and then eliminate them from the current location set.
Then reapply the procedure on the set of remaining locations to obtain the second category, and so on. We end up with five income categories in both Beijing and Shanghai (see Fig. \ref{fig:Lorenz_Curve} and Table \ref{tbl:income_level}). 
Note that the fifth categories can be further divided, but then the number of locations will be too small to produce meaningful collective results.  
Different from other studies that using empirical experience or arbitrary thresholds to make classification, the applied Loubar method is more objective with thresholds derived directly from the data per se.

\begin{table}[] \centering
\caption{Income categories obtained by Loubar method based on house price for all locations in Beijing and Shanghai} \label{tbl:income_level}
\begin{tabular}{ccccccc}
\hline
\multicolumn{1}{l}{} & \multicolumn{3}{c}{Beijing}                                                               & \multicolumn{3}{c}{Shanghai}                                                               \\ \cline{2-7} 
                & \begin{tabular}[c]{@{}c@{}}House Price\\ (10K RMB)\end{tabular} & \#  & \% & \begin{tabular}[c]{@{}c@{}}House Price \\ (10K RMB)\end{tabular} & \#  & \% \\ \hline
1                    & {[}1355, 8750)                                                  & 274          & 7.75   & {[}2400, 21000)                                                  & 106          & 3.26   \\
2                    & {[}552, 1355)                                                   & 1300         & 36.78  & {[}726, 2400)                                                    & 743          & 22.83  \\
3                    & {[}315, 552)                                                    & 1282         & 36.27  & {[}350, 726)                                                     & 1287         & 39.55  \\
4                    & {[}200, 315)                                                    & 515          & 14.57  & {[}220, 350)                                                     & 835          & 25.67  \\
5                    & {[}60, 200)                                                     & 164          & 4.64   & {[}61, 220)                                                      & 283          & 8.7    \\ \hline
\end{tabular} 
\end{table}

\subsection{Radius of Gyration\label{sec:radius_of_gyration}}
Besides travel distance, 
radius of gyration $r_g$ is another important mobility indicator, which depicts the activity range of individuals. The $r_g$ is formulated as
\begin{equation}
    r_g=\sqrt{\frac{1}{N}\sum^N_{i=1}(\vec{r_i}-\vec{r_c})^2},
\end{equation}
where $N$ is the number of visited locations, $\vec{r_i}$ the vector of longitude and latitude of location $i$, and $\vec{r_c}$ the mass center of all visited places. 
A higher $r_g$ implies a wider activity range of the user. 

\subsection{Diversity\label{sec:entropy}}
The diversity of visitations and residence of users
from different categories is one of essential topics in our study. 
The diversity $H(x)$ is quantified by the normalized entropy as
\begin{equation}
    H(x)=-\frac{\sum_{x\in X} -p(x) \log_2 p(x)}{log_2 m},
\end{equation}
where $x$ is the instance value, $p(x)$ is the respective probability, and $X=\{ x_1, x_2,\dots, x_m \}$ is the set of all $m$ instance values. With the $\log_2 m$ as the denominator, we are able to take-off scale effect and compare results with different sizes.

In this work, $x_i$ can be either the fraction of visitation activities or residential population of users in a certain income category to the location $i$, and the set $X$ is the collection of all locations in the same income category, and $p(x_i)=x_i/\sum_{x_i\in X}x_i$. 
In general, the diversity is low when the feature distribution is concentrated in a few locations, and high if the distribution is more even.


\section*{Acknowledgement}
This work receives financial supports from the National Natural Science Foundation of China (Grant No. 61903020), 
Fundamental Research Funds for the Central Universities (Grant No. buctrc201825). 

\section*{Author contributions statement}

R.L. conceived the study and supervised the work, T.G., Y.Y., C.L., and F.S. conducted the experiments, R.L., C.L., T.G., and Z.X. analysed the results, Z.X., T.G., and R.L. wrote the paper. All authors reviewed the manuscript.

\end{document}